 \documentclass{aa}
 \usepackage{graphicx}
 \usepackage{natbib}
 \bibpunct{(}{)}{;}{a}{}{,}
 \newcommand{\oi}{O\,{\sc i}}
 \newcommand{\oifor}{[O\,{\sc i}]}
 \newcommand{\nai}{Na\,{\sc i}}
 
 \newcommand{\cai}{Ca\,{\sc i}}
 \newcommand{\fei}{Fe\,{\sc i}}
 \newcommand{\feii}{Fe\,{\sc ii}}
 \newcommand{\nii}{Ni\,{\sc i}}

 \newcommand{\logo}{$\log{\epsilon_{\rm O}}$}
 \newcommand{\logfe}{$\log{\epsilon_{\rm Fe}}$}
 \newcommand{\logni}{$\log{\epsilon_{\rm Ni}}$}
 
\newcommand{\feh}{[Fe/H]}

 \newcommand{\ofe}{[O/Fe]}

\newcommand{\teff}{$T_{\rm eff}$}
 
 \newcommand{\logg}{$\log{g}$}
 \newcommand{\micro}{$\xi_{\rm turb}$}

 \newcommand{\kms}{km\,s$^{-1}$}

 \newcommand{\marcs}{{\sc marcs}}

\newcommand{\dex}{\,\mathrm{dex}}
 \newcommand{\bd}{\object{BD$-01\degr\,2582$}}
 \newcommand{\cda}{\object{CD$-24\degr\,1782$}}
 \newcommand{\cdb}{\object{CD$-30\degr\,0298$}}
 
\begin{document}
 \title{Oxygen abundances in metal-poor subgiants as determined from \oifor , 
\oi\ and OH lines\thanks{Based on observations collected at the European Southern 
Observatory, Chile (ESO No. 68.D-0546)}}
 \subtitle{}
 
\author{A.E. Garc\'{\i}a P\'erez\inst{1}$^,${\thanks{Student visitor at the European Southern Observatory, Munich, Germany}}
        \and
         M. Asplund\inst{2}
        \and 
        F. Primas\inst{3}
        \and P.E. Nissen\inst{4} 
       \and 
       B. Gustafsson\inst{1}}
 \offprints{aegp@astro.uu.se}
 \institute{
 Department of Astronomy and Space Physics, Box 515 SE-751 20 Uppsala, Sweden
 \and
 Mt. Stromlo Observatory, Cotter Rd., Weston, ACT 2611, Australia
 \and
 European Southern Observatory, Karl-Schwarzschild Str. 2, D-85748
 Garching b. M\"unchen, Germany
 \and
 Department of Physics and Astronomy, University of {{\AA}}rhus,
 DK-8000 {{\AA}}rhus C, Denmark}
 
\date{Received: soon; accepted: quite soon}
 
\authorrunning{A.E. Garc\'{\i}a P\'erez et al.}
 \titlerunning{Oxygen abundances in metal-poor subgiants}
 
\abstract{The debate on the oxygen abundances of metal-poor stars has its 
origin in
 contradictory results obtained using different abundance indicators. To 
achieve a better
 understanding of the problem we have acquired high quality spectra 
with the Ultraviolet and Visual Echelle 
Spectrograph at VLT, with a signal-to-noise of the order of 100 
in the
 near ultraviolet and 500 in the optical and near infrared wavelength 
range. Three 
different oxygen abundance indicators, OH ultraviolet lines around 310.0\,nm, the 
\oifor\ line
 at 630.03\,nm and the \oi\ lines at 777.1-5\,nm were observed in the spectra 
of 13 metal-poor
 subgiants with $-3.0\le\mathrm{[Fe/H]}\le-1.5$. Oxygen abundances were 
obtained from the analysis of these indicators which was carried out assuming local thermodynamic 
equilibrium and plane-parallel model atmospheres. Abundances 
derived from
 \oi\ were corrected for departures from local thermodynamic equilibrium. Stellar parameters were computed 
using \teff-vs-color calibrations based on the infrared flux method and Balmer
 line profiles, Hipparcos 
parallaxes and \ion{Fe}{ii} lines. [O/Fe] values derived from the 
forbidden line
 at 630.03\,nm are consistent with an oxygen/iron ratio that 
varies linearly with [Fe/H] as $\mathrm{[O/Fe]}=-0.09(\pm0.08)\mathrm{[Fe/H]}+0.36(\pm0.15)$. 
Values based on the \oi\ triplet 
are on average $0.19\pm0.22\dex$(s.d.) higher than the values based on the 
forbidden line
 while the agreement between OH ultraviolet lines and the forbidden line is much 
better
 with a mean difference of the order of $-0.09\pm0.25\dex$(s.d.). In general, our 
results follow
 the same trend as previously published results with the exception of the ones 
based on OH ultraviolet lines. In that case our results lie below the values which gave rise to
 the oxygen abundance debate for metal-poor stars.
 }
 \maketitle
 %
 
 \section{Introduction}
 
Oxygen plays a major role in astrophysics through its high cosmic abundance.
 It is predominantly produced and ejected back into the interstellar
 medium in connection with core-collapse supernovae (SNe II).
 By comparing oxygen abundances in stars of different ages
 with those of iron, which is produced
 both in SNe II and in thermo-nuclear supernovae of white dwarfs (SNe Ia),
 one can probe the star formation rate history and initial mass
 function of a galaxy as well as the physics of supernovae
 \citep[e.g.][]{Tinsley80,Wheeler89}.
 Due to the longer time-scale of Fe production in SNe Ia , one can expect
 an oxygen over-abundance relative to iron at low metallicity:
 \ofe\,$> 0$\footnote{The abundance ratios are defined by the customary
 [X/Fe]\,$=\log{(N_{\rm X} / N_{\rm Fe})_*} - {\log{(N_{\rm X} / N_{\rm 
Fe})_\odot}} $.}.
 
While such an oxygen over-abundance in metal-poor stars
 has long been known to exist \citep{Conti67},
 the exact amount is still contested with no unanimous agreement apparently
 in sight.
 Broadly speaking the derived \ofe\ values depend on which oxygen diagnostic
 is employed in the analysis and what type of stars is observed.
 There are four different types of spectral transitions which have
 been utilised for this purpose: the forbidden \oifor\ 630.0 and 636.3\,nm\ 
lines,
 high excitation \oi\ lines, in particular the \oi\ 777.1-5\,nm\ triplet,
 OH A-X electronic lines in the ultraviolet (UV) and OH vibration-rotation 
lines in the infrared (IR).
 The \oi\ and OH lines have mainly
 been used in metal-poor turn-off stars (\teff\,$\ga 6000$\,K)
 while the O abundances in halo giants (\teff\,$\la 5000$\,K) are
 typically based on the \oifor\ lines.
 The \oifor\ lines suggest a quasi-plateau at \ofe\,$\sim +0.5$ for 
\feh\,$\le -1$ \citep[e.g.][]{Barbuy88,Sneden91,Nissen02}.
 This has traditionally been advocated as the correct metallicity trend,
 in particular since the first studies based on the OH lines in the UV gave
 consistent results \citep{Bessell84,Bessell91,Nissen94}.
 More recently, the OH lines in the IR have also been studied in a few
 stars with a similar result, at least when restricting
 to weak, higher excitation OH lines \citep{Balachandran01,Melendez01}.
 Systematically higher \ofe\ values are normally derived when using the \oi\ 
triplet
 \citep[e.g.][]{Abia89,Israelian98,Israelian01,Boesgaard99,Carretta00,Nissen02,Fulbright03},
 sometimes interpreted as a steady
 linear increase in \ofe\ towards lower {\feh}.
 The \oi\ triplet results have often been blamed on departures from local thermodynamic equilibrium (LTE),
 effects of inhomogeneities
 and/or an erroneous \teff -scale in order to reconcile them with the \oifor\ 
results.
 The issue of oxygen abundances in halo stars received much renewed attention 
following
 the surprising results of \citet{Israelian98,Israelian01} and \citet{Boesgaard99}
 who found a near linear trend in \ofe\ with a significant slope of about 
$-0.4$ towards lower
 metallicity from OH lines in the UV in turn-off stars and a couple of 
subgiants.
 
All of the above-mentioned oxygen diagnostics have their advantages and 
disadvantages.
 The \oi\ triplet is susceptible to non-LTE effects (NLTE) \citep[]
[and references therein]{Kiselman01} 
and is, due to the high excitation potential, quite sensitive to the adopted 
{\teff}.
 The \oifor\ line is immune to departures from LTE but is essentially 
undetectable
 in turn-off stars with \feh\,$\la -2$ due to its weakness.
 The OH lines are very sensitive to the temperature structure in the stellar
 atmosphere. Given the much lower temperatures encountered in the optically
 thin layers in 3D hydrodynamical model atmospheres compared with classical
 1D hydrostatic models for metal-poor stars \citep{Asplund99},
 large downward abundance corrections have been flagged for the OH lines
 in turn-off stars. The \oifor\ lines are also affected by such temperature
 inhomogeneities but to a lesser degree than the OH lines \citep{Nissen02},
 while the NLTE line formation of the \oi\ lines can be suspected to
 be dependent on the 3D atmospheric structure as well \citep{Kiselman95,
 Asplund01,Asplund04}.
 Additional complications and confusion stem from the differences and
 uncertainties in the adopted
 \teff\ and \logg , the use of \fei\ or \feii\ lines to derive \feh , possible
 missing continuous opacities in the UV \citep{Balachandran98,Bell01} and
 which solar oxygen abundance the stellar values are referenced to.
 Given the recent large downward revision of the solar O abundance to
 \logo\,$\sim 8.7$ \citep{Allende01,Asplund04}, this
 difference of about 0.2\,dex compared with the previously advocated value
 \citep{Anders89} will directly translate to a corresponding
 {\em increase} in \ofe .
 
Considering all these remaining uncertainties it is perhaps not surprising
 that no consensus has as yet been achieved in terms of the behaviour of \ofe\ 
with \feh .
 It is noteworthy that significant differences between the different
 oxygen indicators only reveal themselves at low metallicities (\feh\,$\la -2$),
 which is observationally very
 challenging due to the weakness of some of the spectral features.
 The fact that the different oxygen diagnostics are normally used in
 different types of stars constitutes a major problem in this regard.
 The only metal-poor stars for which it is possible to employ molecular as well 
as
 forbidden and permitted atomic lines are subgiants but also then 
exceptionally high-quality spectra are required. Until now, only two halo 
subgiants
 (\object{HD\,140283} and \object{BD$+23\degr3130$}, both with \feh\,$\sim -2.4$)
 have been analysed with \oifor , \oi\ and OH UV lines
 \citep{Israelian98,Israelian01,Fulbright99,Cayrel01,Balachandran01,Nissen02}, 
with somewhat ambiguous results.
 Our best hope to finally resolve this important outstanding problem
 is offered by halo subgiants.
 In the present article we present such a study based on very high-quality
 UV and optical spectra for a sample of 13 subgiants with
 $-3.0 \le {\rm [Fe/H]} \le -1.5$.

\section{Observations}
 \label{observations}
 
We identified a sample of halo subgiants based on Hipparcos parallaxes and/or
 Str\"omgren photometry with metallicity $-3.0 \le {\rm [Fe/H]} \le -1.5$. 
From this sample, high-quality spectra were acquired for 13 subgiants using
 the UVES \citep{Dekker00} at ESO's Very Large Telescope (VLT) 8m Kueyen telescope.
 The observations were carried out in service mode between October 2000 and 
March 2001.
 The use of the dichroic capability of UVES enabled simultaneous observations 
of the UV and optical region
 to cover the three most important oxygen abundance diagnostics:
 OH A-X lines around 310.0\,nm , the \oifor\ 630.03\,nm\ line and the \oi\ 
777.1-5\,nm\ lines.
 Due to the relatively low effective temperature of the targets, the observing
 times were largely determined by the requirement to obtain reasonable 
signal-to-noise ($S/N$) in the UV (resulting in $S/N \sim$ 100-150 at 310.0\,nm ) .
 This enabled the total exposure time to be split between two optical settings 
to cover
 both the forbidden and permitted \oi\ lines and still achieve very high
 $S/N$ ($>500$ in the centre of an echelle order at both 630 and 777\,nm, 
although
 as will be described further below the \oi\ triplet falls close to the edge of 
two orders
 in most cases resulting in typical values of $S/N \sim 250$ at the triplet).
 We thus used two different dichroic settings for the red arm of the 
spectrograph (580 and 860\,nm)
 together with the 346\,nm setting for the blue arm.
 In total, the spectra covered 302.4-388.4\,nm and 476.4-1061.0\,nm, with near 
complete
 coverage in these regions.
 In all cases, at least three different exposures were obtained in each setting 
to
 allow efficient removal of cosmic rays.
 
For the blue arm, a projected slit-width of 1.0$\arcsec$ was used while for 
the red arm the width
 was 0.7$\arcsec$. The resulting nominal spectral resolving powers associated 
with these
 slit-widths were 45\,000 and 55\,000, respectively, with 4 pixels per spectral 
resolution element.
 In order to gain $S/N$ in the UV and because of oversampling, a 2x2 binning 
was used when
 reading out the EEV CCD without loosing any spectral resolution.
 
As there are several telluric lines due to in particular H$_{2}$O and O$_2$ 
around 630.03\,nm,
 the \oifor\ line could be blended by telluric features
 depending on the radial velocity of the star and the particular date of the 
observations.
 We tried to avoid such problems by placing time constraints on when the 
observations should
 be carried out in service mode. For this purpose, we obtained radial velocity 
information
 of our targets from SIMBAD and computed the location of the stellar line in 
relation
 to the telluric lines for the relevant observing period.
 The procedure worked very well with one exception in which the radial velocity 
reported by SIMBAD turned out to be different 
from the actual value (given in Table\,\ref{ph}) by about 
20\,\kms. As a result, our spectra of this star (CD\,$-24\degr\,1782$)
 shows the \oifor\ 630.03\,nm\ significantly distorted by telluric absorption,
 unfortunately rendering it useless for an abundance analysis.

The procedure followed in the data reduction was the classical one:
 subtraction of bias, definition of the echelle orders, division by the 
normalised
 flat-field, extraction of the orders, wavelength calibration and
 continuum normalization, all carried out within {\sc{iraf}}.
 The continuum was fitted using cubic spline functions.
 Due to the metal-poor nature of our targets, it was always relatively
 straight-forward to identify wavelength regions without significant line 
contributions
 in the optical to make the continuum well defined. 
The UV region is more crowded even in metal-poor stars but it was still 
possible
 to find sufficiently many
 apparently clean continuum regions in the neighborhood of the OH lines.
 For the UV spectral region, the $S/N$ was modest and hence the
 UVES pipeline reduction worked well, making a special data reduction 
unwarranted.
 
As mentioned above, care was taken when preparing for the observations 
in order to minimise the risk for blending of the \oifor\ line with telluric 
lines.
 Additional precautions in this regard were taken by obtaining
 spectra of rapidly rotating B stars during the same nights and at about
 the same airmass as the programme stars.
 In a few cases when the stellar \oifor\ line was partially blended
 by such telluric absorption, we relied on the {\sc{iraf}} task {\em telluric}, which
 worked well. The task optimises the fit between the 
telluric lines of the programme star and the B star by allowing a relative 
scaling in airmass and a shift in wavelength. The stars in our sample with 
partially blended \oifor\ lines are \object{HD\,4306}, \object{HD\,26169}, 
\object{HD\,45282}
 , \object{HD\,126587}, \object{HD\,128279} and \object{HD\,218857}.
 
Observed stellar radial velocities for the stars were derived from
 the Doppler shifts of three \cai\
 (612.22, 616.22, 643.91\,nm) and one \fei\ (623.07\,nm) lines.
 The resulting heliocentric velocities are given in Table\,\ref{ph}.
 
\begin{table*}
 \begin{center}
 \renewcommand{\tabcolsep}{4.0pt}
 \caption{\label{ph} List of observed stars including coordinates, $uvby\beta$ 
photometry,
 $(V-K)$ colour, Hipparcos parallaxes
 and the measured radial velocity from our observations
 }
 \begin{tabular}{lcrccccccccrc}
 \hline
\hline
 Star & $\alpha(2000)$ & \multicolumn{1}{c}{$\delta(2000)$} & $V$ & $(b-y)$ & 
$m_1$ & $c_1$ & $\beta$ & $(V-K)$ & $\pi$ & $\sigma(\pi)$ 
&\multicolumn{1}{c}{$v_\mathrm{r}$}\\
 & [hr:min:s]&\multicolumn{1}{c}{[$^\circ$:':'']}&[mag]&[mag]&[mag]&[mag]&[mag]&
 [mag]&[mas]&[mas]&\multicolumn{1}{c}{[\kms ]}\\
 \hline
 
\object{HD\,4306}  & 00:45:27.2 &$-09:32:39.8$ & 9.035 & 0.518 & 0.052 & 0.348 
& 2.529 & 2.212 & 4.81 & 1.40 & $-69.8$\\
 \object{HD\,26169} & 04:00:52.4 &$-75:36:11.5$ & 8.797 & 0.523 & 0.051 & 0.328 
& 2.521 & 2.163 & 2.83 & 0.79 & $-35.6$\\
 \object{HD\,27928} & 04:22:55.1 &$-37:15:49.2$ & 9.554 & 0.506 & 0.067 & 0.295 
& 2.523 & 2.106 & 2.36 & 1.16 & $43.3$\\
 \object{HD\,45282} & 06:26:40.8 &$03:25:29.8$ & 8.028 & 0.451 & 0.108 & 0.277 
& 2.544 & 1.939 & 7.34 & 0.96 & $305.7$\\
 \object{HD\,108317} & 12:26:36.8 &$05:18:09.0$ & 8.036 & 0.447 & 0.055 & 0.291 
& 2.548 & 1.883 & 4.53 & 1.06 &$6.6$\\
 \object{HD\,126587} & 14:27:00.4 &$ -22:14:39.0$ & 9.119 & 0.602 & 0.042 & 
0.460 & ..... & 2.451  & 1.40 & 1.44 &$149.6$\\
 \object{HD\,128279} & 14:36:48.5 &$ -29:06:46.6$ & 8.039 & 0.467 & 0.055 & 
0.262 & 2.545 & 1.974  & 5.96 & 1.32 &$-76.0$\\
 \object{HD\,200654} & 21:06:34.7 &$ -49:57:50.3$ & 9.097 & 0.460 & 0.027 & 
0.271 & 2.534 & 1.943  & 3.20 & 1.25 &$-46.1$\\
 \object{HD\,218857} & 23:11:24.6 &$ -16:15:04.0$ & 8.967 & 0.506 & 0.076 & 
0.371 & ..... & 2.094 & 3.51 & 1.42 &$-170.9$\\
 \object{HD\,274939} & 05:33:18.0 &$ -47:56:13.8$ & 9.435 & 0.494 & 0.128 & 
0.300 & 2.526 & 2.136 & 2.88 & 1.06 &$172.0$&\\
 \bd                 & 11:53:37.3 &$ -02:00:36.7$ & 9.595 & 0.493 & 0.097 & 
0.298 & ..... &1.997& 2.98 & 1.35 &$0.8$&\\
 \cda                & 03:38:41.5 &$ -24:02:50.3$ & 9.934 &0.466 & 0.040 & 
0.287 & 2.526 & 1.975 & 4.42 & 1.75 &$101.0$\\
 \cdb                & 00:58:43.9 &$ -30:05:57.7$ & 10.80 & 0.483 & 0.025 & 
0.296 & ..... &1.973  & 1.06 & 2.23 &$27.2$\\
\hline
\end{tabular}
\end{center}
\end{table*}
 
\section{Model atmospheres and stellar parameters}
 
Before proceeding to determining element abundances for our sample, the stellar
 fundamental parameters must be estimated.
 The effective temperature ($T_{\rm{eff}}$), surface gravity ($\log{g}$),
 metallicity ([Fe/H]) and microturbulence (\micro ) are the parameters
 needed to specify standard 1D model atmospheres.
 Such model atmospheres are used to calculate theoretical stellar spectra which 
can be
 compared with observed spectra directly, or indirectly by comparing observed 
and
 computed equivalent widths of spectral lines.
 
\subsection{Model atmospheres}
 
The transport of radiation in the stellar atmospheres needs to be modelled in 
order
 to enable comparison with the observed spectra.
 For the purpose, we have calculated
 theoretical 1D LTE model atmospheres in hydrostatic equilibrium using the
 \marcs\ program \citep{Gustafsson75, Asplund97} 
for the stellar parameters associated with the observed stellar sample.
 These models are based on up-to-date continuous opacities and include
 the effects of line-blanketing through opacity sampling.
 In the program, the convective energy flux is estimated using the 
mixing-length
 theory, assuming a mixing-length parameter $\alpha = 1.5$.
 The models have all been computed using a microturbulence \micro\,$= 
1.0$\,\kms .
 The adopted solar element abundances for the construction of the model 
atmospheres stem from \citet{Grevesse98} with the exception of O which is taken from the
 1D estimate of \citet{Nissen02}: \logo\,$=8.74$.
 The chemical compositions of the models corresponding to our stellar sample are
 then scaled by the stellar metallicity, assuming that all the 
$\alpha$-elements (with carbon not included) were enhanced by $0.4\dex$.
 
The solar oxygen and iron abundances used to estimate \ofe\ and \feh\ were 
derived
 using a solar \marcs\ model atmosphere with the stellar parameters
 \teff\,$=5780$\,K, \logg\,$=4.44$ [cgs] and \feh\,$=0.00$ and
 adopting a microturbulence \micro\,$=1.15$\,\kms\ for the spectral line
 formation calculations. The basic reason for our choice of a theoretical {\sc{marcs}}
 model to represent the solar atmosphere instead of a semiempirical one, such 
as the
 Holweger-Mueller model, was the ambition to make the analysis differential as 
far
 as possible, where systematic errors in models used for the Sun and the 
programme stars hopefully could cancel to a considerable extent.

\subsection{Reddening}
 
The majority of stars in our sample are more distant than 200\,pc
 according to their Hipparcos parallaxes. For these distances interstellar
 extinction could become important and the observed stellar
 magnitudes and stellar colours have to be corrected of reddening.
 Fortunately, most of the stars are found at high galactic latitudes where 
reddening in general is limited. Two different methods have been considered 
for estimating reddening:
 the maps of infrared dust emission by \citet{Schlegel98}
 and the models of visual interstellar extinction by \citet{Hakkila97}.
 In addition, the calibration of the interstellar doublet Na{\sc{i}}
 lines at 589.0 and 589.6\,nm by \citet{Munari97} have been taken into account
 in some cases for comparison purposes.
 \citet{Schlegel98} used the COBE/DIRBE and IRAS/ISSA infrared 
maps of dust emission over the entire sky to estimate the dust column density.
 To proceed from the dust column density
 to reddening they used colours of elliptical galaxies.
 An important aspect of these maps is the possible presence of
 filamentary details. \citet{Hakkila97} calculated
 reddenings based on published results of large-scale visual interstellar 
extinction.
 While the reddening estimates based on the model
 by Hakkila et al. depend on the assumed stellar distances, the estimates from 
the
 IR dust emission maps of Schlegel et al. give the reddening that an
 object will have if it would lie outside the dusty Galaxy.

\begin{figure}
 \resizebox{\hsize}{!}{\rotatebox{0}{\includegraphics{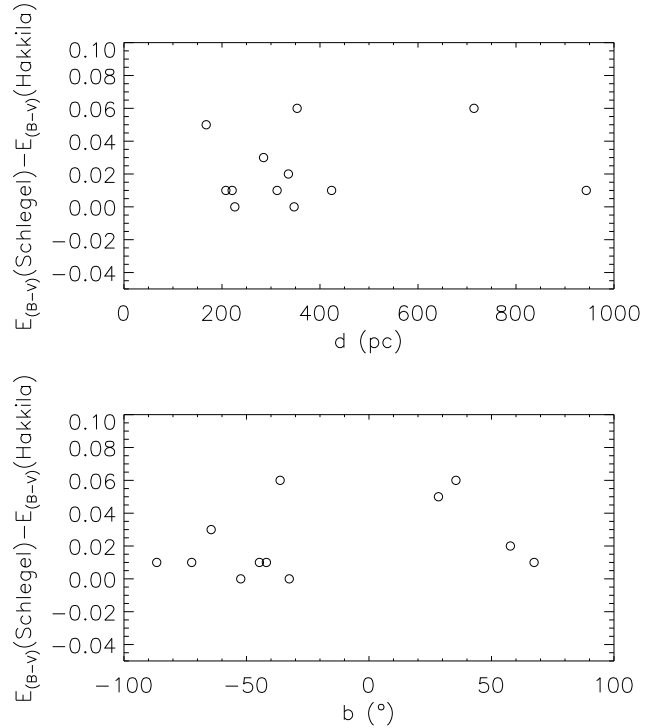}}}
 \caption{\label{figre} Comparison of reddening estimates based on the two 
methods
 by \citet{Hakkila97} and \citet{Schlegel98}. The differences
 are plotted versus the stellar distance in parsecs (d) in the upper panel and
 versus the Galactic latitude of the star in degrees (b) in the bottom panel.
 Note that \object{HD\,45282} is not shown in the figure as it would fall well 
outside
 the plot limits ($\Delta E_\mathrm{(B-V)}=0.78$).
 }
 \end{figure}

\begin{table}
 \begin{center}
 \caption{\label{re} The galactic coordinates and reddenings determined using 
the methods
 of \citet{Hakkila97} and \citet{Schlegel98}}
 \renewcommand{\tabcolsep}{1pt}
 \begin{tabular}{lrrcc}
 \hline
\hline
 Star &\multicolumn{1}{c}{$l$} & \multicolumn{1}{c}{$b$} & $E_{(B-V)}$ & 
$E_{(B-V)}$ \\
    &\multicolumn{1}{c}{[$^\circ$]}&\multicolumn{1}{c}{[$^\circ$]}&[mag]&[mag]\\
 \hline
 \object{HD\,4306}    & 118.057 &$-72.359$   & 0.03 &  0.04 \\
 \object{HD\,26169}   & 289.820 &$-36.268$   & 0.02 &  0.08 \\
 \object{HD\,27928}   & 239.682 &$-44.720$   & 0.02 &  0.03 \\
 \object{HD\,274939}  & 254.431 &$-32.541$   & 0.03 &  0.03 \\
 \object{HD\,45282}   & 207.050 &$-3.930$   & 0.04 &  (0.82) \\
 \object{HD\,108317}  & 286.675 &$67.387$   & 0.01 &  0.02 \\
 \object{HD\,126587}  & 330.345 &$35.489$   & 0.04 &  0.10 \\
 \object{HD\,128279}  & 329.070 &$28.356$   & 0.05 &  0.10 \\
 \object{HD\,200654}  & 348.821 &$-41.873$   & 0.02 &  0.03 \\
 \object{HD\,218857}  &  52.926 &$-64.414$   & 0.01 &  0.04 \\
 \bd                  & 275.101 &$57.703$   & 0.00 &  0.02 \\
 \cda                 & 217.842 &$-52.332$   & 0.02 &  0.02 \\
 \cdb                 & 275.051 &$-86.625$   & 0.01 &  0.02 \\
 \hline
 \end{tabular}
 \end{center}
 \end{table}

The estimated reddening values $E_\mathrm{(B-V)}$ for both the Schlegel et al. 
and Hakkila et al.
 methods are given in Table\,\ref{re} together with the galactic longitude and 
latitude.
 The values based on Schlegel et al. are almost always higher than those derived
 from Hakkila et al.
 In Fig.\,\ref{figre} the differences between the two methods are displayed 
against
 stellar distance and galactic latitude.
 There may be an indication that the greatest differences occur for low 
galactic latitudes.
 It is difficult to assess which values
 are more accurate.
 The uncertainties in the reddening given by Schlegel et al. are claimed to be
 of the order of $16\%$, Hakkila et al. estimate the errors to be about 0.07 
mag.
 Indeed, the differences we have found could be interpreted as an
 indication of the uncertainty in the estimated reddening values. However, we 
believe
 that the Hakkila-based values may give a better description of the interstellar
 dust distribution in the solar neighborhood. We also note that the IR dust 
maps used by
 Schlegel et al. may be contaminated by infrared point sources that could not be
 properly eliminated. One such example is \object{HD\,45282} for which
 the Schlegel et al. value is $E_\mathrm{(B-V)}$=0.82.
 \object{HD\,45282} is a relatively nearby star ($d\,\sim140$\,pc) so the IR 
dust emission maps
 most probably overestimate the reddening, although we note that
 the star is located at low latitudes.
 As will be seen below, with $E_\mathrm{(B-V)}$=0.82 the effective temperature
 for this subgiant would be more than 8000\,K, obviously an impossibly large 
value for a star, e.g. showing OH lines,
 which strengthens our belief that the Hakkila et al. method is to be preferred
 for our programme stars.
 
\begin{figure}
 \begin{center}
 \resizebox{\hsize}{!}{\includegraphics{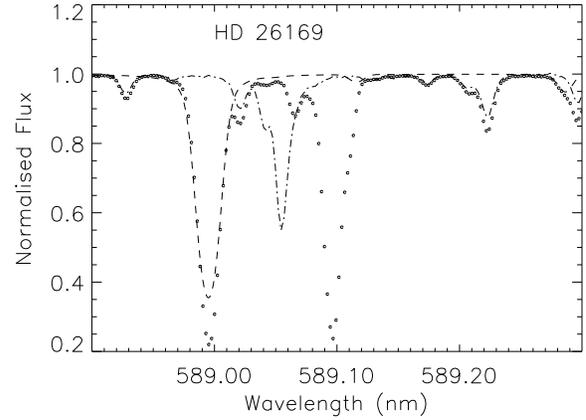}}
 \caption{\label{figna} The observed (open circles) and synthetic (dashed line)
 spectrum of \object{HD\,26169} around the \nai\ D 589.0\,nm\ line. Also shown 
is
 the observed spectrum of the B star \object{HR\,1705} (dot-dashed line), to help 
identifying
 the telluric lines in the observed spectrum of \object{HD\,26169}.
 The spectral feature around 589.1\,nm is the interstellar
 absorption line.}
 \end{center}
 \end{figure}

The equivalent widths of the interstellar \nai\ doublet at 589\,nm
 can also provide an independent estimate of the reddening; our
 VLT-spectra covered this wavelength region as well.
 This has been tried for a few stars for comparison with 
the results of the other two methods. It is not always possible to
 make useful reddening estimates
 with this method due to blending with
 stellar lines or due to multiple interstellar components.
 To isolate the interstellar component of the lines, we performed
 spectrum synthesis using atomic data from the The Vienna Atomic 
Line Data Base (VALD) \citep{Piskunov95}.
 As an example, \object{HD\,26169} shows a relatively strong main
 interstellar absorption as illustrated in Figure \ref{figna}.
 According to \citet{Munari97} the equivalent width of the
 interstellar \nai\ D 589.0\,nm\ absorption line of 15.0\,pm\ corresponds to
 E$_\mathrm{(B-V)} = 0.05$ .
 This is in between the reddening estimates presented in Table\,\ref{re}
 from the Schlegel et al. and Hakkila et al. models. The 
line in the 
case of \object{HD 45282} is much weaker ($\sim2$\,pm) which suggests
 a very low reddening, lower than 0.05. The interstellar contribution to the 
line
 will be even weaker if it turns out to be blended with a telluric line. 
Unfortunately, the interstellar \ion{Na}{i} of the fast-rotator calibration 
star lies
 approximately at the same observer's wavelength as the line of concern. Hence, 
it not possible to identify any telluric blend by just comparing the observed 
spectra
 of these two stars.
 
\subsection{Effective temperature: photometry}

The effective temperatures have been estimated using optical and IR broad-band
 and Str\"omgren photometry, in particular $(V-K)$ and $(b-y)$ colours.
 The $V$ and $uvby\beta$ photometry comes from \citet{Schuster88} and private 
communication, while the $K$ magnitudes come from The Two Micron All Sky Survey (2MASS) 
database. The photometric data together with the parallaxes
 for the programme stars are given in Table\,\ref{ph}.
 As described above, we have used the Hakkila et al. interstellar extinction 
maps
 to estimate reddenings, which have been applied to de-redden the stellar 
photometry.
 To convert the de-reddened $(b-y)$ and $(V-K)$ colours to \teff\ estimates, we have 
used the
 calibrations by \citet{Alonso99}, which are based on the infrared flux method:
 
\begin{eqnarray}
\label{cal1}
\theta_\mathrm{(V-K)} & = & 0.5558+0.2105(V-K)+0.001981(V-K)^2-\nonumber\\
  & & 0.009965(V-K)[Fe/H]+0.01325[Fe/H]-\nonumber\\
  & & 0.002726[Fe/H]^2\\
\theta_\mathrm{(b-y)} & = & 0.5815+0.7263(b-y)+0.06856(b-y)^2-\nonumber\\
  & & 0.06832(b-y)[Fe/H]-0.01062[Fe/H]-\nonumber\\
  & & 0.01079[Fe/H]^2\label{cal2}\\
T_{\rm eff} & = & 5040/\theta \label{cal3}.
 \end{eqnarray}
 
As the subgiants of the observed sample are ascending the red giant branch
 we have used the calibrations for giants rather than for dwarfs. We note, 
however,
 that the differences in temperature between the two calibrations are small.
 In our case, the dwarf calibration would have resulted in about 50\,K lower 
values.
 
\begin{figure}
 \resizebox{\hsize}{!}{\rotatebox{0}{\includegraphics{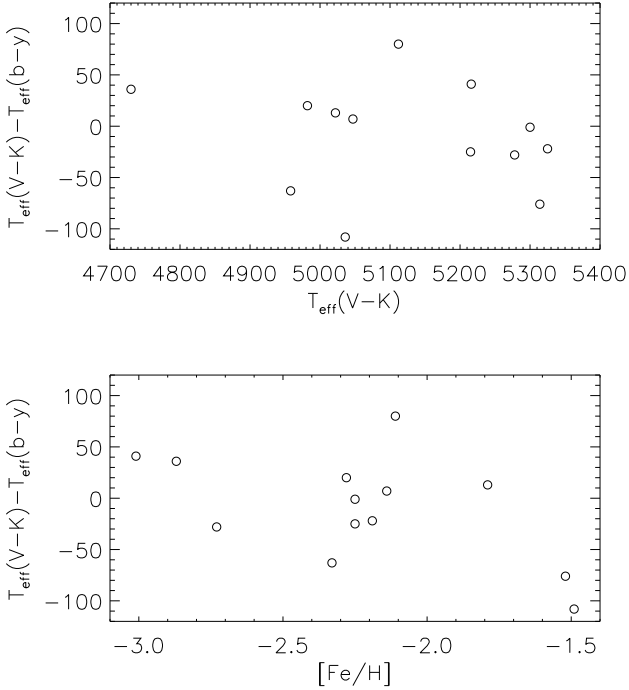}}}
 \caption{\label{figte} Difference in \teff\ when using the de-reddened $(V-K)$ and 
$(b-y)$ colours
 following Eq. \ref{cal1} and Eq. \ref{cal2}, respectively. The temperature 
differences
 are shown in the top panel
 as a function of \teff (determined from $(V-K)$)
 and in the bottom panel versus \feh .}
 \end{figure}

The agreement between \teff $(V-K)$ and \teff $(b-y)$ is quite good, as illustrated
 in Fig.\,\ref{figte}. The mean difference is $9.70 \pm 52$\,K (s.d.) with
 a maximum difference of 108\,K. There is no obvious trend in these differences 
with
 \teff , although there may be a correlation with \feh . The latter conclusion, 
however,
 hinges entirely on the two least metal-poor stars with \feh\,$\simeq -1.5$.
 The mean of values based on the de-reddened $(b-y)_0$ and $(V-K)_0$ colours 
were adopted as the
 final stellar effective temperature. The resulting photometric \teff\ values 
are listed
 in Table\,\ref{sp} together with the mean \teff\ without correcting for 
reddening.
 While accounting for reddening makes a significant difference, the maximum 
correction
 is nevertheless $\le 200$\,K when relying on the \citet{Hakkila97} reddening.
 As expected, the use of the \citet{Schlegel98} method would raise
 the de-reddened \teff\ values greatly in some cases as seen from 
Table\,\ref{sp}.
 As detailed below, those high \teff\ values are in many cases inconsistent with
 those estimated independently from H$\alpha$ line-profile fitting.
 
The typical observational error in $(b-y)$ is 0.007\,mag, which according to Eq. 
\ref{cal2}
 corresponds to an error in \teff\ of about $\pm 35$\,K, assuming an associated
 uncertainty in \feh\ of $\pm 0.1$\,dex.
 The observational error in $(V-K)$ is significantly higher, perhaps as high as 
$\pm 0.05$\,mag,
 but due to the high sensitivity to \teff\ for this colour, this corresponds to 
only $\pm 45$\,K.
 These errors are consistent with the spread in the difference between the two 
colour indicators as shown in
 Fig.\,\ref{figte}.
 The absolute \teff\ values are, however, likely significantly more uncertain 
than this.
 Another important uncertainty stems from the adopted reddening, which
 is hardly better determined than $\Delta E_\mathrm{(B-V)}=\pm 0.02$ ($\Delta 
E_\mathrm{(b-y)}=\pm 0.014$). This error estimate may even seem optimistic in 
view of the large error (0.07 mag.) quoted by \citet{Hakkila97}. However, the 
relatively small distances to our programme stars
 as well as the small reddening values from both \citet{Hakkila97} and 
\citet{Schlegel98} support such a low value.
 This corresponds to an error in \teff\ of about $\pm 70$\,K.
 Furthermore, \teff\ may be affected by other systematic errors, but a 
reasonable estimate of the uncertainty in \teff\ is $\pm 100$\,K.
 
\subsection{Effective temperature: hydrogen lines}
 
The stellar H$\alpha$ 656.3\,nm line is a potentially good indicator of 
effective temperatures and has been used extensively in the literature (e.g. \citet{Fuhrmann93}; 
\citet{Barklem02} and
 references therein).
 The fact that this temperature indicator is not affected by 
reddening is an advantage over photometry-based methods, especially when
 accurate reddening estimates are not available.
 As discussed in detail by \citet{Barklem02} there are, however, several 
problems
 with this methods which must be overcome.
 An important aspect to be considered in the modelling is the theory used to
 describe the broadening of the hydrogen lines, both Stark and self-broadening.
 Additionally, the results are dependent on the employed 1D model atmospheres 
with
 the mixing length parameters used for computing the convective energy flux
 \citep{Fuhrmann93}.
 One may worry that the 1D model atmospheres are not sufficiently realistic
 to describe this aspect. There is also some dependence of \teff\ as derived 
from
 H lines on the adopted stellar gravities.
 A change of 0.3\,dex in $\log{g}$ can affect the estimates of the effective
 temperatures by as much as 60\,K in the case of the metal-poor subgiants.
 Finally, the normalisation of the observed spectra is also a potential source 
of error.
 However, for our subgiants, the wings of the H$\alpha$ line do not extend more 
than one
 spectral order so the normalisation of the spectra, though not trivial is much 
less a
 critical issue than
 when dealing with turn-off stars. Furthermore, given the metal-poor nature of
 our targets, it is straightforward to trace the hydrogen wings between the
 very few other stellar lines in this wavelength region.
 
Our spectra are of sufficiently high quality in terms of resolving power and
 $S/N$ so that the observational errors in the H line profiles
 are insignificant compared with the
 modelling uncertainties (broadening and model atmospheres in particular).
 We estimate that our relative H$\alpha$-based \teff\ values are accurate to 
significantly
 better than 100\,K, although the absolute values could be in error by perhaps 
twice this amount
 \citep[see discussion in][]{Barklem02}.
 As a consequence, the H$\alpha$ method mainly serves here as an independent
 test of the photometric values derived above, which will
 remain our preferred choice for the abundance analysis.
 
\begin{figure*}
 \resizebox{\hsize}{!}{\includegraphics{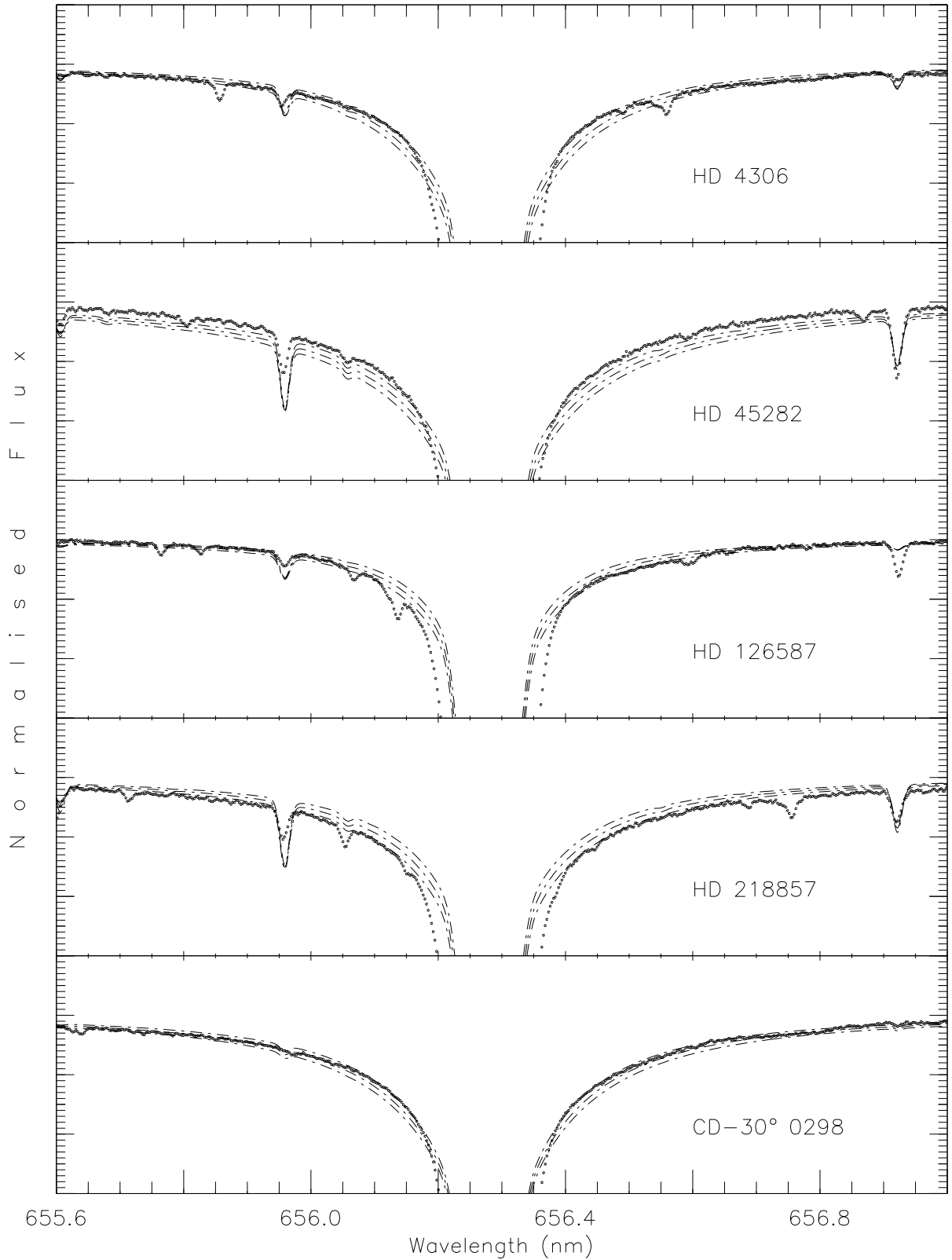}}
 \caption{\label{figha} The observed (open circles) H$\alpha$ line spectra are 
compared
 for five stars with synthetic spectra
 (dashed lines). All spectra have been normalised. Three synthetic spectra are 
plotted for each
 star, corresponding to three different effective temperatures. The middle line 
corresponds to the assumed
 effective temperature based on the de-reddened photometry
 while the other two correspond to 100\,K higher and lower \teff, respectively.
 The scale on the vertical axis goes from 0.7 to 1.1 for each panel.
 The observed absorption lines in the wings of the H$\alpha$ line which do not 
have
 a counterpart in the theoretical profiles are atmospheric H$_2$O lines.
 }
 \end{figure*}

\begin{figure*}
 \resizebox{\hsize}{!}{\includegraphics{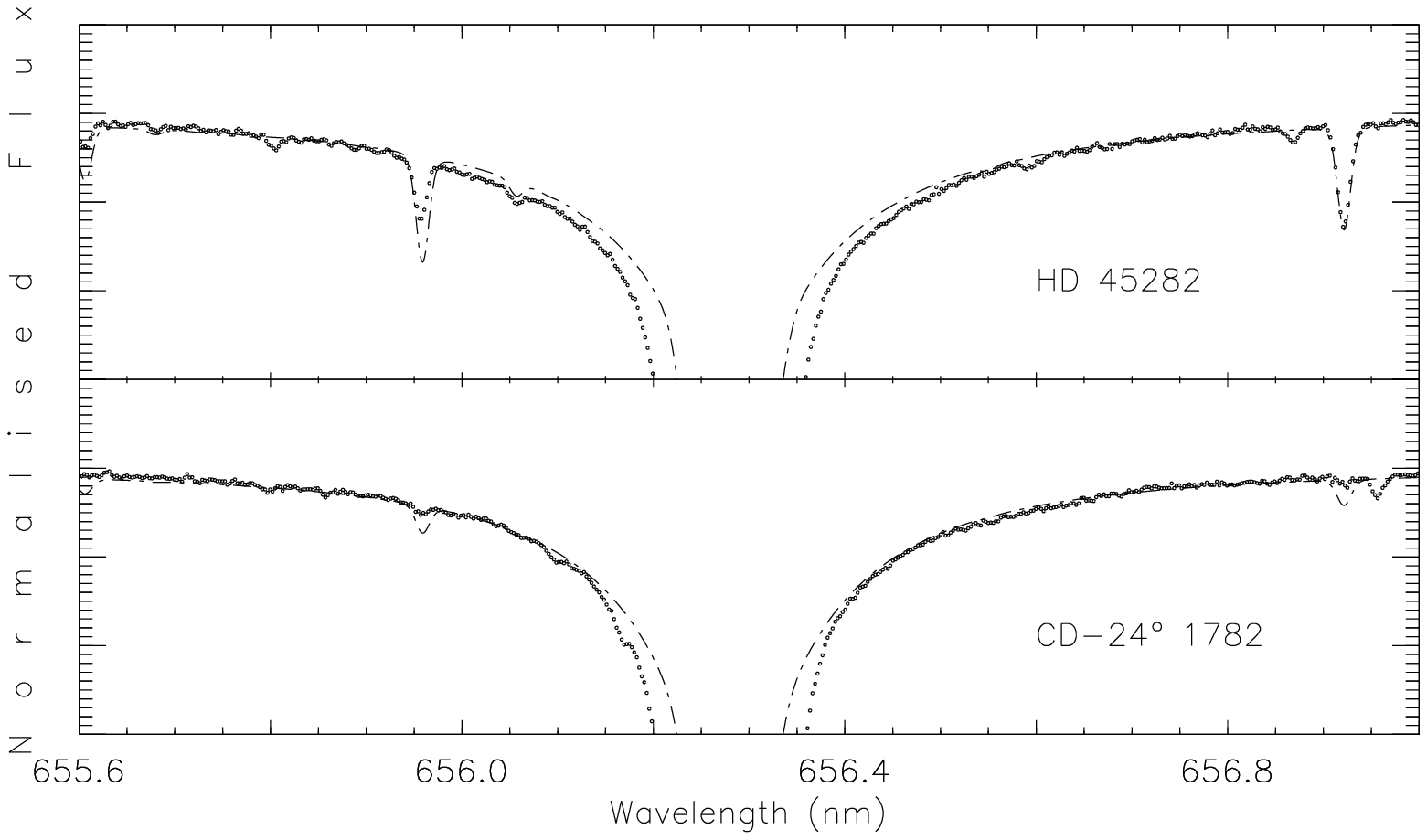}}
 \caption{\label{figha2} The observed H${\alpha}$ line (circles) and the 
best-fitting
 synthetic spectra (dot-dashed lines) for \object{HD\,45282} and \object{CD$-24\degr\,1782$},
 corresponding to \teff\,$=5050$\,K and 4900\,K, respectively.}

\end{figure*}

The observed H${\alpha}$ lines are displayed in Fig.\,\ref{figha} for five 
stars together with
 corresponding synthetic spectra. The theoretical H line profiles shown in the 
figure have
 been computed for the mean de-reddened \teff\ based on $(b-y)$ and $(V-K)$ as well 
as
 for \teff$\pm100$\,K. We note that we only attempt to fit the wings of the H 
line as the core is affected
 by departures from LTE \citep{Przybilla04} and by uncertainties in the 
structure of
 the upper atmosphere and its velocity fields.
 Estimates of effective temperatures based on H$\alpha$ are given in 
Table\,\ref{sp}.
 They agree with the assumed photometric stellar effective temperatures to 
within 100\,K
 except for \object{HD\,45282} and \object{CD$-24\degr\,1782$}. A lower 
effective temperature
 had to be adopted to
 fit their observed H$\alpha$ lines, 5050\,K and 4900\,K versus 5352\,K and 
5228\,K.
 The computed spectra for those new effective temperature values are shown in 
Fig.\,\ref{figha2}.

Estimates of effective temperatures based on H$\alpha$ lines are in this work
 on average about 70 K lower than the values based on photometry associated with
 reddenings from \citet{Hakkila97} and about 10 K higher than values based on 
the non-corrected photometry.
 There is no obvious trend in the differences of H$\alpha$ temperatures and the
 photometric ones with the assumed reddening.
 On the other hand, the values of $T_{\rm{eff}}$ coming from de-reddened 
photometry using
 reddenings from \citet{Schlegel98} are higher by about 150 K than the values 
determined from H$\alpha$.
 This is another indication that the Schlegel et al. dust maps overestimate the 
reddenings for our
 programme stars.
 
\subsection{Gravities}
 
The stellar surface gravities were determined from Hipparcos parallaxes
 as long as the relative error $\Delta \pi / \pi$ in the parallax satisfied 
$\Delta \pi / \pi \le 0.5$.
 This was fulfilled in all cases except for \object{HD\,126587} and 
\object{CD$-30\degr\,0298$}.
 To convert from parallaxes to surface gravities we relied on the fundamental
 relation \citep[e.g.][]{Nissen97}:
 
\begin{eqnarray}
 \label{gr}
  \log{\frac{g}{g\odot}} & = &
  \log{\frac{M}{M_\odot}}+4\log{\frac{T_{\mathrm{eff}}}{T_{\mathrm{eff\odot}}}}+
 \nonumber\\
 & & 0.4V_{0}+0.4BC+2\log{\pi}+0.12.
 \end{eqnarray}
 
The stellar mass $M$ was assumed to be $0.8\,M_\odot$ for all stars which, 
however, is
 not a critical assumption for this method. The bolometric corrections $BC$ 
were taken from \citet{Bessell98}.
 The estimates of the surface gravities based on Eq.\,\ref{gr} are presented in
 Table\,\ref{sp}. The main source of error for the stellar gravities based on 
this method
 is in most cases the adopted parallaxes.
 We estimate the total uncertainty in surface gravity by adding in quadrature 
the
 errors due to the quoted errors in Hipparcos parallaxes (see Table\,\ref{ph}),
 an uncertainty of 100\,K in \teff\  and 0.05\,$M_\odot$ in mass.
 These 1$\sigma$ errors are also listed in Table\,\ref{sp}.
 We note that the effects of uncertain reddening are small as they affect
 \teff\ and visual magnitude in opposite directions and therefore mainly cancel 
each other.
 The typical uncertainty in \logg\ is $\pm (0.2-0.4)$.

An independent method of deriving surface gravities relies on the use of 
theoretical
 evolutionary isochrones for the assumed mass. We have used this method as a 
test, and
 it is our main method
 for the two stars \object{HD\,126587} and \object{CD$-30\degr\,0298$} without 
reliable Hipparcos parallaxes.
 We used the isochrones of \citet{Bergbusch01}, which
 are based on the evolutionary tracks of \citet{VandenBerg00}.
 Interpolation of the isochrones in $\alpha$-enhancement and in metallicity was
 performed, adopting 0.4 and 12\,Gy for the $\alpha$-enhancement and age, 
respectively,
 for all the stars. As corresponding theoretical isochrones are not available 
for \feh\,$\le -2.31$,
 the isochrone calculated for the lowest metallicity 
was adopted for \object{HD\,126587} and \object{CD$-30\degr\,0298$} which are 
both
 quite metal poor, as well as for \object{HD\,200654} ($\mathrm{[Fe/H]} = -2.75$).
 The rest of our programme stars have metallicities of about $-2.3$ or higher.
 
\begin{figure}
  \resizebox{\hsize}{!}{\includegraphics{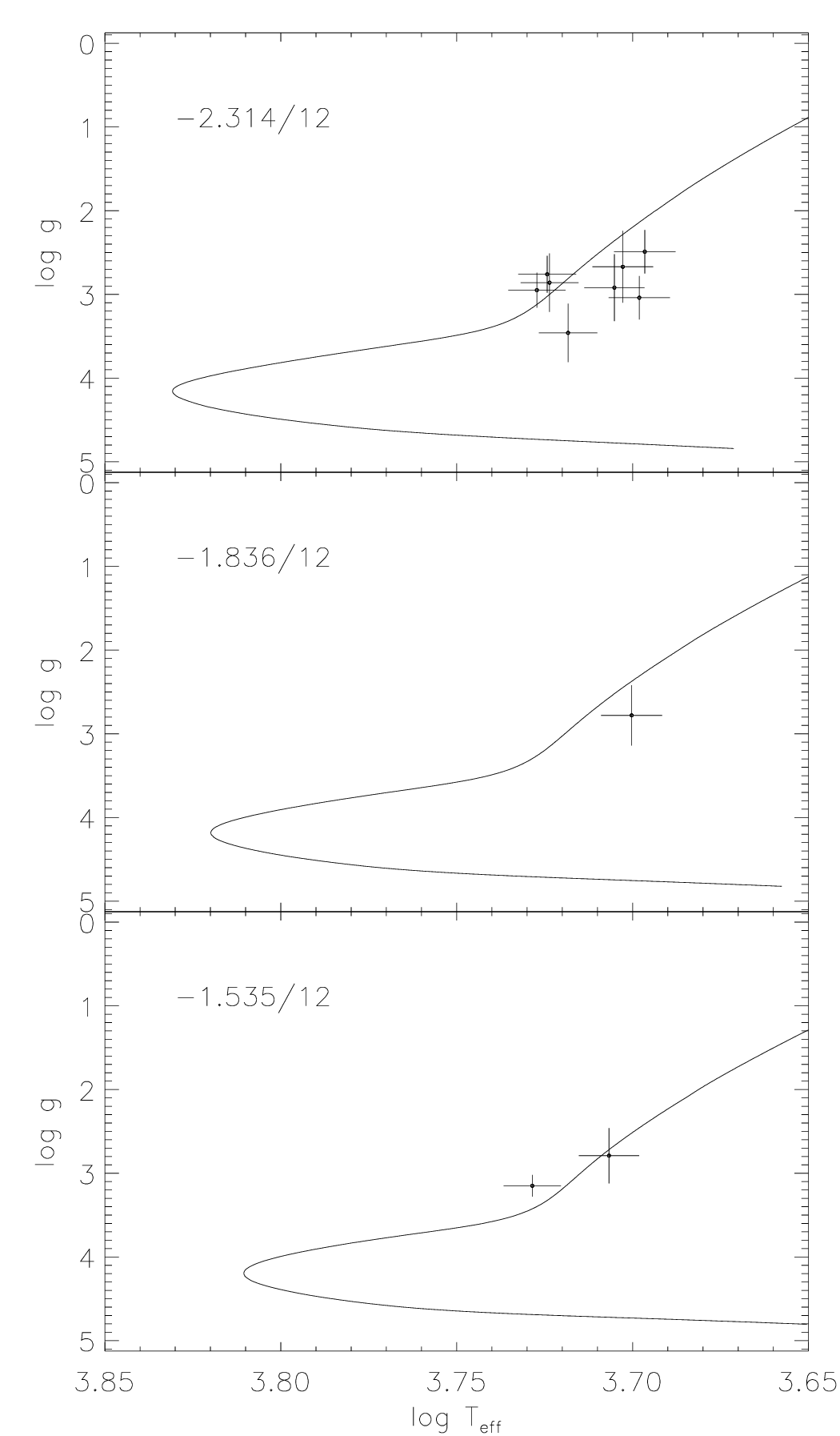}}
 \caption{\label{iso} Comparison of the stellar gravities from Hipparcos 
parallaxes with
 isochrones from \citet{Bergbusch01}. The three panels display the results
 for three difference metallicities: \feh\,$=-2.31$, $-1.84$ and $-1.54$. In 
each case
 the isochrone at an age of 12\,Gyr has been used. The stellar data in the
 [$\log{g},\,\log{T_\mathrm{eff}}$] plane 
are represented by circles with crosses denoting the estimated errors.
 The stars are distributed between the panels according to their
 metallicities.}
  \end{figure}

The resulting surface gravities based on isochrone fitting are listed in 
Table\,\ref{sp}.
 Fig.\,\ref{iso} shows a comparison of the isochrones with the surface 
gravities estimated
 from Hipparcos parallaxes.
 As can be seen, the agreement with the stars for which accurate parallaxes
 are available are quite good.
 The mean difference in \logg\ amounts to $0.21 \pm 0.13$.
 We have used these differences to correct the isochrone-based values for
 \object{HD\,126587} and \object{CD$-30\degr\,0298$} to place them on the same 
scale as the other stars which are based on Hipparcos parallaxes.
 The mean difference for the six stars with \feh\,$\le -2.2$ is 
$\Delta$\logg\,$=0.22 \pm 0.50$(s.d.). The final
 \logg\ values for these two stars listed in Table\,\ref{sp} have 
thus
 been corrected by 0.22. The error in logarithmic gravity for these stars is 
estimated to be
 $\pm 0.5$. Fortunately, the method 
seems to be rather insensitive to metallicity.
 
\begin{table*}
 \renewcommand{\tabcolsep}{2pt}
 \begin{center}
 
\caption{\label{sp} The derived stellar parameters of our sample.
 The effective temperatures determined from reddening corrected colour indexes,
 $(b-y)_0$ and $(V-K)_0$, with reddening values according to column 4 in 
Table\,\ref{re}.
 The means of those effective temperatures are presented in the
 fourth column denoted by $T_{\rm{eff}}$ (adopted). In the fifth column, 
denoted by $T_{\rm{eff}}$ (no-red),
 the same type of mean values are reported but with
 no reddening corrections applied. The mean values for the effective 
temperatures based on reddenings from \citet{Schlegel98} are
 listed in the sixth column. In the seventh column 
values for effective temperature based on the H$\alpha$ line profiles are 
listed.
 The remaining columns give values adopted for the model atmospheres for 
gravities (based on
 Hipparcos parallaxes and isochrones, respectively), metallicities and 
microturbulence parameters.}

\begin{tabular}{lcccccccccc}
 \hline
\hline
 $Star$ & $T_{\rm{eff}}(b-y)$ &$T_{\rm{eff}}(V-K)$ & $T_{\rm{eff}}$ & 
$T_{\rm{eff}}$ & $T_{\rm{eff}}$ & $T_{\rm{eff}}$(H$\alpha$) & \multicolumn 
{2}{c}{$\log{g}$} & [Fe/H] &
 $\xi_\mathrm{turb}$ \\
 &Hakkila et al.&Hakkila et al.&adopted&no-red&Schlegel et al.&& 
Hipparcos&Isochrones &  &\\
 &[K]&[K]&[K]&[K]&[K]&[K]&[cgs]&[cgs] &[dex]&[\kms ]\\
 \hline
 
\object{HD\,4306}   & 5021 & 4958 & 4990 & 4890 & 5024 & 4940 & 3.04$\pm$0.26 & 2.14 &$-2.35$ & 1.5\\
 \object{HD\,26169}   & 4962 & 4982 & 4972 & 4906 & 5181 & 4972 & 2.49$\pm$0.26 & 2.10 &$-2.29$ & 1.5\\
 \object{HD\,27928}  & 5040 & 5047 & 5044 & 4976 & 5078 & 5044 &  2.67$\pm$0.43 & 2.33  &$-2.16$ & 1.5\\
 \object{HD\,45282}  & 5390 & 5314 & 5352 & 5208 & 8170 & 5050  & 3.15$\pm$0.13 & 3.44  &$-1.53$ & 1.1\\
 \object{HD\,108317} & 5301 & 5300 & 5300 & 5263 & 5338 & 5300 &  2.76$\pm$0.22 & 3.05 &$-2.27$ & 1.5\\
 \object{HD\,126587} & 4694 & 4730 & 4712 & 4590 & 4907 & 4812 & (1.66$\pm$0.50) & 1.44 &$-2.88$ & 1.5\\
 \object{HD\,128279} & 5347 & 5325 & 5336 & 5151 & 5533 & 5236 &  2.95$\pm$0.21 & 3.16 &$-2.21$ & 1.5\\
 \object{HD\,200654} & 5306 & 5278 & 5292 & 5216 & 5331 & 5192 & 2.86$\pm$0.35 & 3.01 &$-2.75$ & 1.5\\
 \object{HD\,218857} & 5009 & 5022 & 5015 & 4982 & 5117 & 4915 & 2.78$\pm$0.36 & 2.39 &$-1.80$ & 1.4\\
 \object{HD\,274939} & 5144 & 5036 & 5090 & 4991 & 5090 & 5040  & 2.79$\pm$0.33 & 2.73 &$-1.51$ & 1.3\\
 \bd          & 5032 & 5112 & 5072 & 5072 & 5142 & 5022  & 2.92$\pm$0.40 & 2.41 &$-2.13$ & 1.5\\
 \cda         & 5240 & 5215 & 5228 & 5155 & 5228 & 4900  & 3.46$\pm$0.35 & 2.82 &$-2.27$ & 1.5\\
 \cdb         & 5175 & 5216 & 5196 & 5158 & 5234 & 5044  & (2.93$\pm$0.50) & 2.71 &$-3.04$ & 1.5\\
\hline
 \end{tabular}
 \end{center}
 \end{table*}
 
\subsection{Metallicity and microturbulence}\label{metallicity}
 
The procedure to determine spectroscopic stellar parameters must be iterative.
 The effective-temperature and surface-gravity estimates
 depend on the adopted stellar metallicity, which
 in turn depends on the other two parameters.
 We used the Str\"omgren $m_1$ index to obtain a starting guess for the 
metallicity
 of the model atmosphere by
 which we derived new \feh\ values using observed \feii\ lines.
 We have employed the same 13  \feii\ lines as in \citet{Nissen02} with the 
same adopted
 line properties.
 These new \feh\ values were subsequently used to derive new \teff\ and \logg\ 
values for the stars.
 New \marcs\ model atmospheres were computed with these new parameters from 
which revised
 \feh\ estimates were calculated. The procedure was repeated until convergence
 which occurred within a few iterations.
\begin{table}
\begin{center}
\caption{\label{fe} The derived \feh\ values based on \feii\ lines with 
line-to-line scatter given. Also listed are the
 associated errors in \feh\ due to uncertainties in stellar parameters
 ($100$\,K for \teff , $ 0.1$\,dex in \feh\ and the errors in \logg\ as 
given in Table\,\ref{sp}). Finally, the impact of a measurement error of 0.05\,pm\ for 
the \feii\ equivalent widths on the derived Fe abundances is also illustrated.}
\renewcommand{\tabcolsep}{1.0pt}
\begin{tabular}{lcccccc}
\hline
\hline
Star & [Fe/H] &\multicolumn{5}{c}{$\Delta\mathrm{[Fe/H]}$}\\
& & $\Delta T_{\rm{eff}}$ & $\Delta \log g$ & $\Delta$[Fe/H] & $\Delta W_\lambda$ & $-\Delta W_\lambda$ \\
\hline
\object{HD\,4306}   &$-2.33\pm0.07$ &$ -0.01$ &  0.10 &  0.01 &  0.07 & $-0.10$ \\
\object{HD\,26169}  &$-2.28\pm0.04$ &$  0.00$ &  0.10 &  0.01 &  0.06 & $-0.06$ \\
\object{HD\,27928}  &$-2.14\pm0.04$ &$ -0.01$ &  0.17 &  0.01 &  0.04 & $-0.06$ \\
\object{HD\,45282}  &$-1.52\pm0.04$ &$  0.00$ &  0.05 &  0.02 &  0.03 & $-0.02$ \\
\object{HD\,108317} &$-2.25\pm0.04$ &$  0.00$ &  0.07 &  0.00 &  0.05 & $-0.08$ \\
\object{HD\,126587} &$-2.87\pm0.05$ &$  0.00$ &  0.18 &  0.00 &  0.07 & $-0.10$\\
\object{HD\,128279} &$-2.19\pm0.08$ &$  0.00$ &  0.07 &  0.00 &  0.06 & $-0.09$ \\
\object{HD\,200654} &$-2.73\pm0.08$ &$  0.01$ &  0.12 &  0.00 &  0.08 & $-0.11$ \\
\object{HD\,218857} &$-1.79\pm0.04$ &$ -0.02$ &  0.14 &  0.02 &  0.03 & $-0.03$ \\
\object{HD\,274939} &$-1.49\pm0.04$ &$ -0.02$ &  0.13 &  0.02 &  0.02 & $-0.02$ \\
\bd\                &$-2.11\pm0.07$ & $-0.01$ &  0.15 &  0.01 &  0.05 & $-0.06$ \\ 
\cda\               &$-2.25\pm0.08$ & $-0.01$ &  0.14 &  0.00 &  0.09 & $-0.10$ \\
\cdb\               &$-3.01\pm0.10$ &  0.01 &  0.19 &  0.00 &  0.10 & $-0.14$ \\
\hline
\end{tabular}
\end{center}
\end{table}
The spectroscopic metallicity corresponding
 to the last iteration was adopted as the final stellar metallicity (Table\,\ref{fe}). 
We note that these values slightly differ from those listed in Table\,\ref{sp}, which correspond 
to the assumed model atmosphere metallicity for the last iteration. 

It should be noted that while knowledge about the stellar microturbulence is 
necessary
 in this iterative procedure, the exact choice for this parameter is not 
important
 for the majority of our stars.
 Only for the three most metal-rich stars in the sample (\object{HD\,274939}, 
\object{HD\,218857} and \object{HD\,45282})
 did the derived \feii -based abundances depend on the microturbulence.
 The microturbulence parameters for those stars were estimated by requiring 
that the \feii\ lines
 should not yield any trend in Fe abundance with line strength.
 The thus estimated microturbulence parameters were in the range 
$1.1-1.4$\,\kms .
 For the other stars, the lines are sufficiently weak to be insensitive to the 
choice
 of microturbulence, which were assumed to be 1.5\,\kms . The model atmospheres 
are hardly
 affected at all by the choice of the parameter.
 
The equivalent widths of the \feii\ lines were measured with the {\sc{iraf}} 
task {\em{splot}}.
 Due to the weakness of the lines at low metallicity, all 13 \feii\ lines
 were not detected in all stars. In the case of \object{CD$-30\degr\,0298$} 
there were only 4 lines
 detected while at least 11 lines were detected in the spectra of the rest of 
the stars.
 The choice of ionized iron lines was natural since these lines are
 less affected by departures from LTE (such as over-ionization)
 than neutral iron lines, even if
 the magnitude of the NLTE effects for \fei\ lines is still debated,
 mainly as a result of the uncertain inelastic H collisions (compare for 
example the
 contrasting results of \citet{Gratton99,Thevenin99,Korn03}).
 The \feii\ lines of our programme stars do not suffer
 from any apparent blends \citep{Nissen02}.
 Iron abundances for each of these lines were derived from the measured 
equivalent widths
 adopting the $gf$-values of \citet{Nissen02} which are based on the work by
 \citet{Biemont91}.
 The mean abundance was taken as the stellar metallicity.
 Since we also performed a corresponding solar analysis using the same lines to 
derive
 a solar Fe abundance, the effects of any errors in the $gf$-values on the 
stellar
 iron abundances relative to that of the sun are minimised.

\begin{table}
\begin{center}
 \caption{\label{fei} List of \ion{Fe}{i} lines used for iron abundances calculations, excitational 
potential and $\log{gf}$ values from \citet{Obrian91}.} 
 \begin{tabular}{ccc}
\hline
\hline
$\lambda$ & $\chi$ & $\log{gf}$ \\
{[nm]}&[eV]\\
 \hline
   613.66 &     2.45  &  -1.41 \\    
   613.77 &     2.59  &  -1.35 \\   
   619.16 &     2.43  &  -1.42 \\    
   623.07 &     2.56  &  -1.28 \\    
   625.26 &     2.40  &  -1.77 \\    
   639.36 &     2.43  &  -1.58 \\    
   641.17 &     3.65  &  -0.72 \\    
   642.14 &     2.28  &  -2.01 \\    
   643.09 &     2.18  &  -1.95 \\    
 \hline
 \end{tabular}
 \end{center}
 \end{table}

We have estimated the sensitivity of the derived metallicities to the adopted
 stellar parameters by repeating the analysis with perturbed values for \teff , 
\logg , \feh\ and
 \micro .
 For this exercise we have used our estimated uncertainties in the various 
parameters:
 $100$\,K for \teff , $0.1$\,dex in \feh\ and the quoted errors in \logg\ 
(from Hipparcos parallaxes) given in
 Table\,\ref{sp}.
 The changes in the  \feh\ values in each of these cases are listed in 
Table\,\ref{fe}.
 The dependence of temperature on the line and continuum opacity is such that 
the \feii -based
 abundances are essentially immune to errors in \teff ($\Delta {\rm log} 
\epsilon_{\rm Fe} \le 0.02$\,dex).
 Similarly, the resulting impact on the model atmosphere structure by slight 
changes in metallicity
 is very small for these metal-poor stars and has virtually no effect on the 
derived Fe abundances.
 However, the uncertainties in the stellar gravities lead to significant errors 
in
 the calculated iron abundances. In the worst cases this error may be as large 
as 0.19 dex
 for the stars with poorly determined parallaxes, although in most cases the 
uncertainty is
 about 0.10\,dex.
 The microturbulence influences only the iron abundance slightly even for the 
most metal-rich stars.
 Even a drastic change such as going from 1.1\,\kms\ to 1.7\,\kms\ in the case 
of
 \object{HD\,45282} makes a difference of a mere 0.04\,dex in  \feh .
 
Another source of error for the Fe abundances is the measured equivalent 
widths,
 in particular given the weakness of some of the employed lines in the most 
metal-poor stars.
 Based on photon statistics and the measured $S/N$ around the \feii\ lines we 
estimate
 that the uncertainties in the equivalent widths are at most about 0.05\,pm. 
New Fe abundances with such errors added to the measured line strengths were 
derived
 and are also listed in Table\,\ref{fe}.
 Fortunately, in all cases such uncertainties in the measured equivalent widths 
only
 correspond to errors in the derived mean \feii\ abundances of $\le 0.14$\,dex 
even in the
 unrealistic case when all \feii\ lines are affected fully and in the same 
direction.

\begin{table}
\begin{center}
 \caption{\label{feiab} Stellar iron abundances determined from the \ion{Fe}{i} lines in Table\,\ref{fei}. 
Last column reports \ion{Fe}{i}-\ion{Fe}{ii} iron abundance differences.}
 \begin{tabular}{lcc}
\hline
\hline
$Star$ & $\log{\epsilon}_{\mathrm{Fe}}$ & $\Delta\log{\epsilon}_{\mathrm{Fe}}$\\
&& (\ion{Fe}{i}-\ion{Fe}{ii})\\
\hline
\object{HD\,4306}   &$4.65$ &$-0.56$\\
\object{HD\,26169}  &$5.04$ &$-0.23$\\
\object{HD\,27928}  &$5.17$ &$-0.24$\\
\object{HD\,45282}  &$6.30$ &$0.26$\\
\object{HD\,108317} &$5.24$ &$-0.05$\\
\object{HD\,126587} &$4.41$ &$-0.26$\\
\object{HD\,128279} &$5.39$ &$0.04$\\
\object{HD\,200654} &$4.69$ &$-0.11$\\
\object{HD\,218857} &$5.53$ &$-0.23$\\
\object{HD\,274939} &$5.99$ &$-0.07$\\
\bd\                &$5.12$ &$-0.32$\\
\cda\               &$4.84$ &$-0.46$\\
\cdb\               &$4.22$ &$-0.30$\\
\hline
\end{tabular}
\end{center}
\end{table}

In the literature and so often, one can find iron abundances for 
metal-poor solar type stars derived from neutral iron lines rather than 
from ionised lines. The neutral lines are easier to detect 
because they are in general stronger than the ionised lines. However, these 
lines may suffer from departures from LTE in the line formation 
(\citet{Gratton99,Thevenin99,Asplund05,Collet05}) so we would expect abundances 
derived from them to be different from the iron abundances derived from 
\ion{Fe}{ii} lines. We investigated this by deriving iron abundances from the 
measured equivalent widths of a set of \ion{Fe}{i} lines 
(see Table\,\ref{fei}). A trend may exist for abundance differences between \ion{Fe}{i} 
and \ion{Fe}{ii} lines with the assumed metallicity; lower metallicities seem to show larger 
differences. On average, our assumed metallicity values are 0.2\,dex higher 
than the values derived from the neutral lines (see Table\,\ref{feiab}), except for 
two stars (\object{HD\,45282} and \object{HD\,128279}). Furthermore, we note that two 
of the most metal-poor stars of our sample show the largest differences 
(\object{HD\,4306} and \cda). The fact that \ion{Fe}{ii}-based metallicities are lower 
than the ones derived from \ion{Fe}{i} lines is not easy to 
explain in terms of the NLTE effects because \ion{Fe}{i} lines should give lower values 
than \ion{Fe}{ii} lines. 

Let's now take a closer look at those objects sticking out of the general trend, i.e. the two stars for which 
the iron abundances derived from \ion{Fe}{ii} lines are lower than those derived from \ion{Fe}{i} lines, 
and the two metal-poor stars with the largest \ion{Fe}{i}$-$\ion{Fe}{ii} differences. The first possible 
source of the observed discrepancy are the stellar parameters, which could have been wrongly determined for 
these four stars. In the case of \object{HD\,45282} and \object{HD\,128279}, we note they have the 
highest \teff\ of the entire sample, as well as the largest assumed reddening. This may clearly indicate 
that our effective temperatures for these two stars have been overestimated. Indeed, a decrease of 100~K in 
their temperatures (at the border-line of our quoted uncertainty on this stellar parameter) would bring 
\ion{Fe}{i} and \ion{Fe}{ii} abundances in satisfactory agreement (within 0.1~dex), although with opposite 
sign. For the latter two stars, \object{HD\,4306} and \cda, we note that their Hipparcos-based gravities 
are much higher than the values derived for stars of similar stellar characteristics. Again, this could 
be a hint that our gravity estimates for these two stars are not the optimal values, which could well 
explain the very large differences observed between \ion{Fe}{i}- and \ion{Fe}{ii}-based abundances (up 
to 0.5~dex!). If we were to lower their gravity values to, e.g., 2.5 and 3.0 respectively (from 3.04 and 
3.46, but still within a two sigma error), we would then derive a \ion{Fe}{i}$-$\ion{Fe}{ii} abundance 
difference of the order of 0.3\,dex which is in more accordance with the average value we find (0.2\,dex). 
Note that the isochrones-based gravity values for both stars are even lower than 2.5 and 3.0, which would 
lead to an even better agreement between \ion{Fe}{i} and \ion{Fe}{ii}.

\section{Oxygen abundances}
 
In general it is reasonable to assume that the more different types of 
abundance indicators
 are being used, the
 more reliable are the results, provided that the various transitions yield 
consistent results.
 There are four different indicators for stellar oxygen abundances, three of 
which are
 located in the wavelength range covered by UVES. 
 In essentially all previous work in the extensive literature on oxygen 
abundances in metal-poor stars, the abundances have been derived from only one or in some cases
 two of these indicators simultaneously.
 We have chosen to target subgiants and giants near the base of the red giant 
branch at exceptionally
 high $S/N$ both in the UV and optical regions with the aim to exploit the
 unique opportunity to derive abundances
 from three basically different abundance indicators.
 
In this section we describe how the oxygen abundances have been derived for 
the different
 types of lines. In the case of the forbidden [O\,{\sc i}] and the 
O\,{\sc i} IR triplet lines we have employed both
 the techniques of measuring equivalent widths and detailed spectrum synthesis, 
while
 only the latter is feasible for the crowded spectrum around the OH UV lines.
 The equivalent widths have been measured using the {\sc{iraf}} task called 
{\em{splot}},
 either by directly integrating the area of the spectral line of interest or by 
fitting
 the spectral lines with a Gaussian or another type of function and then 
integrating that curve.
 As the $S/N$ of our spectra is superb, the direct integration method is in 
general 
preferred, since a Gaussian can not perfectly describe a stellar spectral line 
due to the line asymmetries induced by the convective motions in the stellar 
atmosphere.
 In a few occasions a Gaussian fit had to be adopted due to presence of 
telluric lines
 disturbing the profile of the [O\,{\sc i}] line.
 
\begin{figure}
 \begin{center}
 \resizebox{\hsize}{!}{\includegraphics{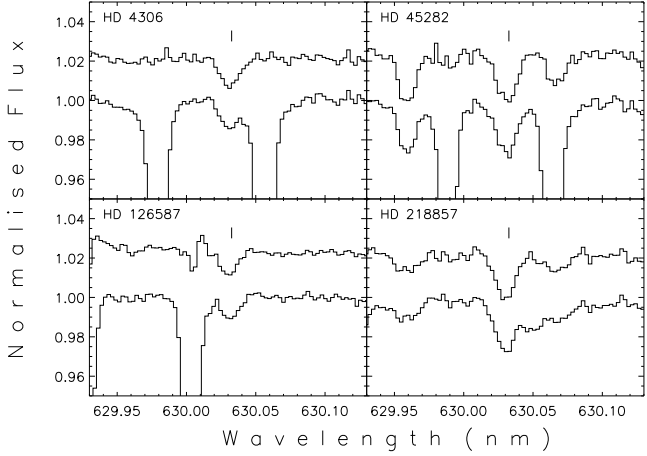}}
 \caption{\label{oiob} Observed spectra with the forbidden 
[O\,{\sc i}] line at 630.03\,nm marked with a vertical bar for 
four of the programme stars. The appearance of the spectra before and 
after correction for telluric absorption is shown;  
 the corrected spectra were re-scaled along the vertical scale.}
 \end{center}
 \end{figure}
 
\subsection{\label{sun} Solar abundances}
 
In order to obtain accurate [O/Fe] values for the stars, we
 must also derive consistent solar O and Fe abundances.
 The solar iron abundance was determined from each of the 13 \feii\ lines used 
to
 derive the stellar Fe abundances with transition probabilities 
and equivalent widths from \citet{Nissen02}, measured in the solar flux atlas 
of \citet{Kurucz84}.
 The solar analysis proceeded completely analogously to the stellar Fe 
determinations.
 The thus derived solar Fe abundance is \logfe\,$=7.56 
\pm 0.09$ (standard deviation).
 The absolute abundance is reasonably close to the quoted meteoritic abundances 
of
 7.50 in \citet{Grevesse98} and 7.45 in \citet{Asplund05}.
 Indeed, had the oscillator strengths from VALD been adopted for these lines,
 solar iron abundance would have decreased to 7.44 while the line-to-line 
dispersion
 diminished to 0.05\,dex.
 The relatively high scatter among the different lines is similar to that
 achieved in the 3D analysis of solar \feii\ lines by \citet{Asplund00b},
 and presumably mainly reflects errors in the oscillator strengths.
 Minor sources of error are inaccurate equivalent widths and 
departures from LTE in the line formation.
 Unlike for our metal-poor stars, in the
 solar case the \feii\ lines of interest are sufficiently strong (2.0-9.0\,pm\ 
with
 9 lines $<4.5$\,pm )
 that errors in line broadening may also play a role.
 If the Uns\"old enhancement factor for the van der Waal's broadening were 
changed from
 1.5 to 2.0 the solar iron abundance would only change by 0.015\,dex. Much more 
important
 is the impact of the microturbulence: the metallicity changed by $-0.16\dex$ 
when the
 assumed value of microturbulence was altered from 1.15 to 2.00\,\kms. By 
computing
 \feh\ values for each individual line for the stars in our sample,
 we removed much of the uncertainty in the $gf$-values and could thus achieve a 
considerably
 reduced
 line-to-line scatter ($\le 0.05$\,dex for seven of our stars and only in one
 case as high as 0.10\,dex).
 
A solar O abundance was determined from the \oifor\ 630.03\,nm line using the
 measured equivalent width of 0.55\,pm\ from \citet{Nissen02}. The 636.38\,nm 
line is more
 severely blended and was not used.
 As also the \oifor\ 630.03\,nm line is blended, with a Ni\,{\sc i} line 
\citep{Allende01},
 we have subtracted its contribution from the feature by adopting a solar
 Ni abundance of \logni\,$=6.25$ \citep{Grevesse98} and
 $\log{gf}=-2.31$ \citep{Allende01}.
 The remaining equivalent width of 0.41\,pm attributable to the \oifor\ 
630.03\,nm line
 implies a solar O abundance of \logo\,$=8.74$. That value was always taken as 
the
 solar O abundance independently of the type of indicator used when deriving 
[O/Fe] for
 the programme stars.

 
\subsection{Oxygen abundances from the \oifor\ 630.03\,nm line}
 
Oxygen abundances were derived from the \oifor\ 630.03\,nm line using both 
measured equivalent
 widths and by means of spectrum synthesis.
 The measured equivalent widths are given in Table\,\ref{oi}.
 Depending on the radial velocity of the star the stellar line can be blended 
by telluric
 lines which have to be carefully considered, in particular for these 
metal-poor stars for which
 the stellar line is so weak. As mentioned in Sect.\,\ref{observations}, 
precautions were taken to
 avoid such blends and only in one case, for \object{CD$-24\degr\,1782$}, the 
\oifor\ line
 turned out to be so severely blended that it was useless for abundance 
determinations.
 
In six additional cases telluric lines partially blended the oxygen feature. 
In the spectra
 of these six stars we used the {\sc{iraf}} task {\em{telluric}}. 
However, since the $S/N$ values for these B stars were lower than for the 
programme stars, this
 method degrades the $S/N$, and consequently should be avoided if possible. The 
procedure worked very well
 for \object{HD\,4306}, \object{HD\,26169}, \object{HD\,218857}  and 
\object{HD\,45282}, but less
 satisfactorily for \object{HD\,126587} and \object{HD\,128279}. The residuals 
(i.e. the difference
 between the observed and the calculated spectra, normalised to the continuum) 
for
 these two latter stars were not insignificant, being of the order of 1-2\%, so 
equivalent widths were not
  measured from them. Instead, a Gaussian profile was fitted to the observed 
lines to estimate
  the contribution of the blend to the wings. In the case of 
\object{HD\,126587} and contrary to the
 case of \object{HD\,128279}, the large residuals are due to a large difference 
in airmass between the star and the corresponding calibration star, which makes the telluric 
correction more uncertain.
 Another reason was that the observations were taken at different times during 
the night for
 the programme star and for its comparison star. 
It should be noted that in the successful cases, blending by telluric lines is 
not particularly
 problematic and a reliable equivalent width measurement could be obtained even 
without
 the {\em telluric} procedure. This can be seen in Fig.\,\ref{oiob} where a few 
spectra before
 and after applying the task are plotted. In the remaining stars, there was no 
need to
 perform any division with a B star spectrum as no telluric lines were 
disturbing the forbidden
 oxygen line. For those spectra, the equivalent widths were measured directly, using direct integration of the 
area of the feature
 in the observed spectrum. In the case of \object{CD$-30\degr\,0298$} and 
\object{HD\,200654} no convincing line detection could be 
made and only an upper limit to the measured equivalent width of $0.04$\,pm\ 
and $0.08$\,pm could be placed. The last value is based on a 
tentative measurement of the equivalent width. 
 
As is obvious from Table\,\ref{oi}, the measured equivalent widths are very 
small, in particular
 at the lowest \feh. It is therefore paramount to carefully estimate the 
uncertainties in
 these values. From the typical $S/N$ of 500 around the \oifor\ 630.03\,nm line, we
 expect an observational error of $\le 0.03$\,pm .
 This calculated value is based on the assumption of a Poisson distribution for 
the signal,
 considering noise only in the line and not in the continuum.
 Fortunately, for these metal-poor stars the placement of the continuum is 
straightforward and
 does not add significantly to the error.
 This was demonstrated by performing repeated independent measurements of the 
equivalent widths
 by different people. In all cases the rms of these measurements were $\le 
0.02$\,pm.

\begin{table}
 \begin{center}
 \caption{\label{oi} The measured equivalent widths of the \oifor\ 630.03\,nm 
line
 and the resulting 1D LTE oxygen abundances. The third column gives the oxygen 
abundances from
 the measured equivalent widths. The remaining columns describe the
 errors in the derived abundances resulting from uncertainties in
 \teff , \logg , \feh\ and of a conservative
 measurement error of 0.05\,pm ;
 the stellar parameter changes are the same as in Table\,\ref{fe}.
 No value is given for \object{CD$-24\degr\,1782$} since the forbidden line is 
badly blended by
 a telluric line. Only an upper limit could be obtained for \object{CD$-30\degr\,0298$} and \object{HD\,200654}.}
\renewcommand{\tabcolsep}{1.0pt}
 \begin{tabular}{lccccccc}
\hline
\hline
$Star$ &$W_{\lambda}$&{$\log{\epsilon}_{\mathrm{O}}$}& \multicolumn{5}{c}{$\Delta\log{\epsilon}_{\mathrm{O}}$} \\
 &[pm]& &$\Delta T_{\rm{eff}}$ & $\Delta \log g$ & $\Delta$[Fe/H] & $\Delta{W_\lambda}$ &$-\Delta{W_\lambda}$ \\
 \hline
 
\object{HD\,4306} & 0.25 & 7.14 &  0.05 & 0.10 & 0.02 & 0.08 & $-0.10$ \\ 
\object{HD\,26169}& 0.33 & 7.05 &  0.06 & 0.10 & 0.02 & 0.07 & $-0.07$ \\ 
\object{HD\,27928}& 0.25 & 7.06 &  0.05 & 0.16 & 0.02 & 0.08 & $-0.10$ \\ 
\object{HD\,45282}& 0.41 & 7.72 &  0.04 & 0.05 & 0.02 & 0.05 & $-0.06$ \\ 
\object{HD\,108317}& 0.25 & 7.23 &  0.07 & 0.07 & 0.01 & 0.08 &$-0.10$ \\ 
\object{HD\,126587}& 0.20 & 6.32 &  0.08 & 0.17 & 0.01 & 0.10 &$-0.12$ \\ 
\object{HD\,128279}& 0.11 & 6.96 &  0.07 & 0.07 & 0.01 & 0.17 &$-0.25$ \\ 
\object{HD\,200654}& $<0.08$ & $<6.75$ &...&...&...&...&...\\        
\object{HD\,218857}& 0.37 & 7.33 &  0.04 & 0.15 & 0.03 & 0.06 &$-0.06$ \\
\object{HD\,274939}& 0.74 & 7.76 &  0.01 & 0.11 & 0.01 & 0.03 &$-0.03$ \\ 
\bd & 0.25 & 7.17 &  0.05 & 0.15 & 0.02 & 0.08 & $-0.10$ \\ 
\cda &... &...&... &...&...&...&...\\
 \cdb & ${<0.04}$ & {$<6.38$} &... &...&...&...&...\\
 \hline
 \end{tabular}
 \end{center}
 \end{table}
 
The \oifor\ line is blended with a \nii\ line at 630.04\,nm \citep{Allende01}.
 The theoretical equivalent widths for this line were computed for the stellar 
parameters in
 Table\,\ref{sp}, adopting Ni abundances as the solar abundance \logni\,$=6.25$
 scaled to the stellar metallicity. We have adopted the $gf$-value for the 
\nii\ line given by
 \citet{Allende01}. For the most metal-rich stars of our sample, the
 Ni contribution is only 0.01\,pm\ and much smaller for the remaining stars.
 Consequently, it is safe to assume that the entire measured equivalent width 
given in
 Table\,\ref{oi} is due to oxygen.
 
The measured equivalent widths have been converted to
 stellar oxygen abundances by using the LTE spectral line formation code {\sc 
eqwidth}
 of the Uppsala spectrum synthesis package, and {\sc marcs} theoretical 1D-LTE 
model atmospheres.
 The $gf$-value for the \oifor\ line was taken as log\,$gf = -9.72$, which is 
the same as
 adopted by \citet{Allende01} and \citet{Nissen02} in their analyses of the 
line.
 The line is in all cases sufficiently weak that uncertainties in the van der 
Waal's pressure
 broadening, radiative broadening and microturbulence are completely negligible.
 The resulting oxygen abundances are given in Table\,\ref{oi}.
 
We have investigated the sensitivity of the derived results to the 
uncertainties
 in the stellar parameters. Table\,\ref{oi} gives the oxygen abundances when 
perturbing
 the stellar parameters by their typical errors ($\pm 100$\,K in \teff , the 
quoted
 uncertainties in \logg\ in Table\,\ref{sp} and $\pm 0.1$ in \feh ).
 The highest sensitivity is for \logg\ where the errors are typically 0.10\,dex 
but
 can be as large as 0.17\,dex. The corresponding errors for \teff\ 
uncertainties are
 about 0.05\,dex while the derived oxygen abundance is insensitive to the
 employed \feh .
 One of the major advantages with using \feii\ lines to estimate \feh\ is that
 the sensitivity of the \feii\ and \oifor\ lines to the stellar parameters is 
very similar.
 Indeed, the thus derived \ofe\ values are practically independent of the 
choice of \logg\ ($\le 0.01$\,dex)
 and only moderately sensitive to \teff\ ($\sim 0.05$\,dex).
 It is therefore clear that our final error budget for \ofe\ as obtained from
 the \oifor\ line will be dominated by the measurement error.
 Since the abundance error will be dependent on $\Delta W_\lambda/W_\lambda$, 
the
 estimated (conservative) observational error of 0.05\,pm\ will have the 
largest impact for
 the most metal-poor stars in our sample.
 
Spectrum synthesis was also applied with the aim of
 checking the oxygen abundances determined from the measured equivalent widths.
 This was done using the spectrum synthesis code {\sc bsyn} which is also a 
part of the Uppsala package.
 The necessary line broadening due to macroturbulence was determined using
 nearby \feii\ lines. The agreement between the equivalent width and spectrum 
synthesis values
 is excellent. The equivalent width-based abundances were finally adopted.
 
\citet{Kiselman93} has studied the NLTE line formation of the forbidden 
oxygen line. According to his
 calculations, the LTE approximation is perfectly justified. In cool stars such 
as ours, oxygen is
 mostly in the ground level of the neutral state because
 of its high ionization energy and the high energies of the excited atomic 
levels.
 The forbidden line is a transition between the ground state and the first 
excited level of \oi ,
 which are closely collisionally coupled. Departures from LTE are therefore 
insignificant for this line.

\subsection{Oxygen abundances from the \oi\ 777.1-5\,nm lines}
 
\begin{figure}
 \begin{center}
 \resizebox{\hsize}{!}{\includegraphics{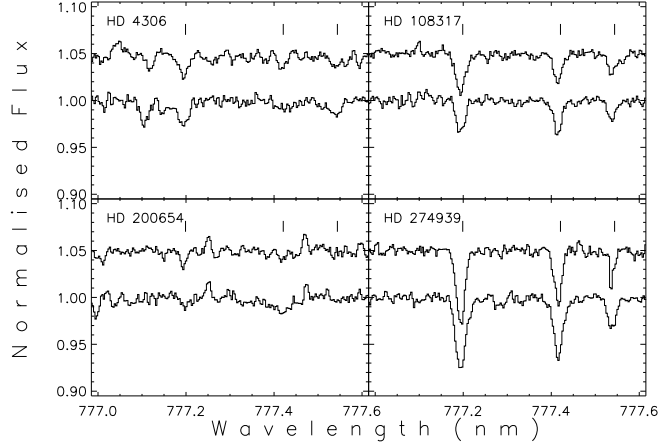}}
 \caption{\label{iroiob} Two adjacent spectral orders containing the \oi\ triplet at 777.1-5\,nm,
 scaled to similar continuum levels. The positions of the triplet lines are indicated by vertical
 bars.}
 \end{center}
 \end{figure}

The \oi\ triplet at 777.1-5\,nm falls near the edges of two adjacent spectral 
orders in the UVES
 standard configurations that we had to use due to the service mode nature of 
our observational
 programme.
 Therefore, while the $S/N$ is excellent at wavelengths near the centre of the 
orders,
 the counts decrease rapidly towards the location of the triplet.
 Typically we measure $S/N$ of $\ge 250$ around the lines but there can be
 substantial variations for a given order of a star
 even within the 0.5\,nm wavelength range of the three triplet lines.
 An additional complication is residual fringing which could not be removed 
fully after
 flat-fielding. 
 
As for the \oifor\ line, O abundances were estimated from equivalent widths.
The oscillator strengths for the transitions were adopted from 
\citet{Wiese96}: $\log{gf}=0.369$, 0.223 and 0.002 for the
 777.19\,nm, 777.42\,nm and 777.54\,nm lines, respectively.
 For completeness we note that we used the Uns\"old formula with an enhancement 
factor of 2.5
 for the van der Waal's pressure broadening of the lines. This has no impact on
 the derived abundances given the weakness of the lines. Likewise, even a 
change in
 microturbulence by 1\,\kms\ introduces a change in O abundance from the 
triplet by $\le 0.02$\,dex.

The equivalent widths of the three triplet lines were measured in both echelle 
orders
 for each star. In a few cases, no reliable equivalent width could be measured 
due to
 insufficient $S/N$ or peculiar profiles. The values from the two orders
 were averaged, weighted with the 
$S/N$ of the spectra around the lines. The quality of the spectra in the 
different spectral orders is illustrated in Fig.\,\ref{iroiob}. Based on the 
comparison of the two sets of equivalent widths, we estimate that
 the error in the mean equivalent width is $\le 0.1$\,pm. 

From these equivalent widths, oxygen abundances were derived for each line, 
which were
 subsequently averaged to yield the \oi\ triplet-based abundance for the stars.
 The equivalent widths and the thus derived O abundances are given in 
Table\,\ref{iroi}.
 The three lines normally give consistent abundances with typical line-to-line 
scatter
 of 0.08\,dex. In one case the largest and smallest values differ by 0.22\,dex. 

\begin{table*}
 \begin{center}
 \caption{\label{iroi} The mean and s.d. of the equivalent widths of the 
\oi\ 777.1-5\,nm triplet lines measured from two different spectral orders 
and the derived oxygen abundances.
 The abundances given in columns 5-7 are the equivalent width-based LTE 
abundances for the three lines
 with the mean for those and the s.d. listed in column 8. The last column give the 
abundances corrected for
 departures from LTE (see text).}
 \renewcommand{\tabcolsep}{4pt}
 \begin{tabular}{lcccccccccc}
 \hline
 \hline
 $Star$ & \multicolumn{3}{c}{$W_{\lambda}$}&& \multicolumn{5}{c}{$\log{\epsilon}_{\mathrm{O}}$}\\ 
\cline{2-4} \cline{6-10}\\
  &$777.1$\,nm &$777.4$\,nm &$777.5$\,nm &&$777.1$\,nm &$777.4$\,nm &$777.5$\,nm &  & NLTE\\
 & {[pm]} & [pm] & [pm]\\
\hline
 \object{HD\,4306}   & $0.63\pm0.16$ &   $ 0.24        $ & $0.39       $ & &  $7.46\pm  0.13 $ &   $7.16        $ &   $ 7.60         $ & $7.40\pm0.22$ &7.33   &\\
 \object{HD\,26169}  & $0.81\pm0.08$ &   $ 0.53\pm0.10 $ & $0.42\pm0.11$ & &  $7.39\pm  0.05 $ &   $7.32\pm0.09 $ &   $ 7.43\pm 0.14 $ & $7.38\pm0.05$ & 7.31&\\
 \object{HD\,27928}  & $0.67\pm0.06$ &   $ 0.58\pm0.05 $ & $0.28\pm0.08$ & &  $7.28\pm  0.05 $ &   $7.36\pm0.04 $ &   $ 7.22\pm 0.13 $ & $7.29\pm0.07$ & 7.22&\\
 \object{HD\,45282}  & $2.06\pm0.08$ &   $ 1.77\pm0.23 $ & $1.17\pm0.14$ & &  $7.79\pm  0.03 $ &   $7.84\pm0.09 $ &   $ 7.81\pm 0.07 $ & $7.82\pm0.03$ & 7.70&\\
 \object{HD\,108317} & $0.94\pm0.13$ &   $ 0.69\pm0.07 $ & $0.50\pm0.06$ & &  $7.24\pm  0.07 $ &   $7.23\pm0.06 $ &   $ 7.29\pm 0.06 $ & $7.25\pm0.03$ & 7.17&\\
 \object{HD\,126587} & $0.46       $ &   $ 0.35        $ & $0.15       $ & &  $7.09          $ &   $7.10        $ &   $ 6.94         $ & $7.04\pm0.09$ & 6.97&\\
 \object{HD\,128279} & $1.07\pm0.33$ &   $ 0.62\pm0.00 $ & $0.39\pm0.11$ & &  $7.34\pm  0.17 $ &   $7.21\pm0.00 $ &   $ 7.20\pm 0.13 $ & $7.25\pm0.08$ & 7.17&\\
 \object{HD\,200654} & $0.31\pm0.05$ &   $ 0.18\pm0.02 $ & $ ...       $ & &  $6.74\pm  0.07 $ &   $6.66\pm0.05 $ &   $ ...          $ & $6.70\pm0.06$ & 6.63&\\
 \object{HD\,218857} & $1.05\pm0.13$ &   $ 1.00\pm0.12 $ & $0.57\pm0.09$ & &  $7.60\pm  0.07 $ &   $7.71\pm0.07 $ &   $ 7.63\pm 0.09 $ & $7.64\pm0.06$ & 7.56&\\
 \object{HD\,274939} & $2.01\pm0.35$ &   $ 1.39\pm0.40 $ & $0.62\pm0.23$ & &  $7.91\pm  0.13 $ &   $7.81\pm0.18 $ &   $ 7.58\pm 0.20 $ & $7.77\pm0.17$ & 7.67&\\
 \bd                 & $1.03\pm0.09$ &   $ 0.88\pm0.02 $ & $0.47\pm0.04$ & &  $7.57\pm  0.05 $ &   $7.64\pm0.02 $ &   $ 7.54\pm 0.03 $ & $7.58\pm0.05$ & 7.50&\\
 \cda                & $0.51\pm0.10$ &   $ 0.35\pm0.06 $ & $...         $ & &  $7.27\pm  0.09 $ &   $7.25\pm0.07 $ &   $ ...          $ & $7.26\pm0.01$ & 7.18\\
 \cdb                & $<0.2$&$...         $ & $...         $ && $<6.68$  &$ ...        $ &   $ ...          $ & $<6.68$&6.61 \\
 \hline
 \end{tabular}
 \end{center}
 \end{table*}
 
We have investigated the uncertainty in the derived oxygen abundances by
 varying the stellar parameters and the equivalent widths in the same fashion as
 for the forbidden oxygen line. The results of this exercise are
 shown in Table\,\ref{senoir}. Contrary to the abundances based on the 
forbidden oxygen line, those based on 
the triplet lines decrease when the effective temperatures increase due to
 the high excitation potential of the levels of these lines.
 The population of the excited levels increases with temperature
 such that an increase of 100\,K in \teff\ results in a decrease of 0.1\,dex in 
abundance.
 As oxygen is predominantly neutral in the conditions prevailing in the 
atmospheres of our stars,
 the \oi\ lines are sensitive to the gravity.
 Just like for the \oifor\ line, the abundance sensitivity is about 0.1\,dex 
for the
 uncertainties of our \logg\ estimates.
 The sensitivity to \logg\ for the triplet is fortunately largely offset
 by the similar effect on the \feii\ lines when estimating \ofe . 

The triplet is insensitive to the metallicity of the stellar atmosphere model.
 Since the triplet is quite weak for the relatively low \teff\ of our stellar 
sample,
 in particular for the most metal-poor stars, the derived abundances are 
sensitive to
 the errors in the measured equivalent widths. Our errors of
 about 0.1\,pm lead to an abundance error of as much as 0.26\,dex
 for the weakest line in the most metal-poor stars, but more typically 
0.05-0.10\,dex. 
 
It has since long been established that the \oi\ triplet is affected by
 departures from LTE in late-type stars \citep[e.g.][and references therein]{Kiselman01}.
 We have estimated the NLTE effects on the derived O abundances by performing
 specifically tailored statistical-equilibrium calculations for our stars, in an
 identical fashion to those by \citet{Nissen02}.
 We have used a 23 level model atom with 43 bound-bound transition and with all 
22 \oi\ levels
 connected to O\,{\sc ii} by photo-ionization transitions.
 This is sufficiently extensive to yield reliable results since the line 
formation
 of the triplet is anyway largely an equivalent two-level problem.
 The radiative data comes mainly from the Opacity Project \citep{Cunto93}.
 Collisional excitation and ionization with electrons and charge transfer 
reactions
 are accounted for but inelastic collisions with H have been neglected (see 
discussion
 in \citet{Kiselman01}).
 The NLTE calculations have been performed using {\sc multi} version 2.2 
\citep{Carlsson86} for three different abundances surrounding the LTE based abundances derived 
from
 the triplet lines for the stars. From these calculations, NLTE abundance 
corrections
 were interpolated and added to the LTE results in Table\,\ref{iroi}.
 For our stars, the NLTE abundance corrections typically amount to about 
$-0.08$\,dex.

\begin{table}
 \begin{center}
 \renewcommand{\tabcolsep}{3pt}
 \caption{\label{senoir} The sensitivity of the oxygen abundances from the \oi\ 
triplet lines
 to uncertainties in the stellar parameters and equivalent widths. The stellar 
parameter changes are the same as in Table\,\ref{fe}.}
 \begin{tabular}{lccccc}
 \hline
 \hline
$Star$ &\multicolumn{5}{c}{$\Delta\log{\epsilon}_{\mathrm{O}}$} \\
   &$\Delta T_{\rm{eff}}$ & $\Delta \log g$ & $\Delta$[Fe/H] & 
$+\Delta{W_\lambda}$ & $-\Delta{W_\lambda}$\\
 &\multicolumn{4}{c}{LTE}  \\
 \hline
 \object{HD\,4306}   & $-0.11$ &  0.11 & $-0.01$ &  0.09 & $-0.13$ \\
 \object{HD\,26169}  & $-0.10$ &  0.10 &  0.00 &  0.08 & $-0.09$ \\
 \object{HD\,27928}  & $-0.10$ &  0.17 & $-0.01$ &  0.09 & $-0.12$ \\
 \object{HD\,45282}  & $-0.10$ &  0.04 & $-0.01$ &  0.02 & $-0.03$ \\
 \object{HD\,108317} & $-0.09$ &  0.08 &  0.00 &  0.06 & $-0.07$ \\
 \object{HD\,126587} & $-0.11$ &  0.22 &  0.00 &  0.14 & $-0.24$ \\
 \object{HD\,128279} & $-0.09$ &  0.08 &  0.00 &  0.07 & $-0.09$ \\
 \object{HD\,200654} & $-0.09$ &  0.14 &  0.00 &  0.16 & $-0.26$ \\
 \object{HD\,218857} & $-0.10$ &  0.13 &  0.00 &  0.05 & $-0.06$ \\
 \object{HD\,274939} & $-0.11$ &  0.12 &  0.00 &  0.04 & $-0.05$ \\
 \bd                 & $-0.10$ &  0.16 &  0.00 &  0.06 & $-0.06$ \\
 \cda                & $-0.10$ &  0.14 & $-0.01$ &  0.10 & $-0.12$ \\
 \cdb                &  ...  &  ...  &  ...  &  ...  &  ...  \\
 
\hline
 \end{tabular}
 \end{center}
 \end{table}


 
\subsection{Oxygen abundances from the OH UV lines}
 
The high quality of our observed spectra in the UV-blue wavelength region has 
also
 enabled the use of the OH (A-X) electronic lines around 310\,nm for abundance 
determination.
 One of the main problems with these diagnostics is the crowded spectral region
 they are located in. This necessitates the use of spectral synthesis rather
 than equivalent widths for deriving oxygen abundances.
 The high $S/N$ and resolving power of the blue spectra also allow an
 accurate placing of the continuum which 
is a non-trivial task for this region.
 Furthermore, it makes it easier to distinguish possible blends.
 That problem is the least for the most metal-poor stars in our sample where 
the continuum
 is relatively easy to locate and blends less disturbing, while the
 most metal-rich stars are likely to have larger abundance uncertainties due 
to this.
 An initial normalisation of the UVES pipeline reduced spectra was done by 
fitting
 the upper envelope of the spectra to a cubic spline function in {\sc{iraf}}.
 A minor re-normalisation around the OH lines was subsequently performed in 
some cases
 when the comparisons between the normalised observed and synthetic spectra 
clearly
 required it.
 
\begin{figure*}
 \begin{center}
 \resizebox{\hsize}{!}{\includegraphics{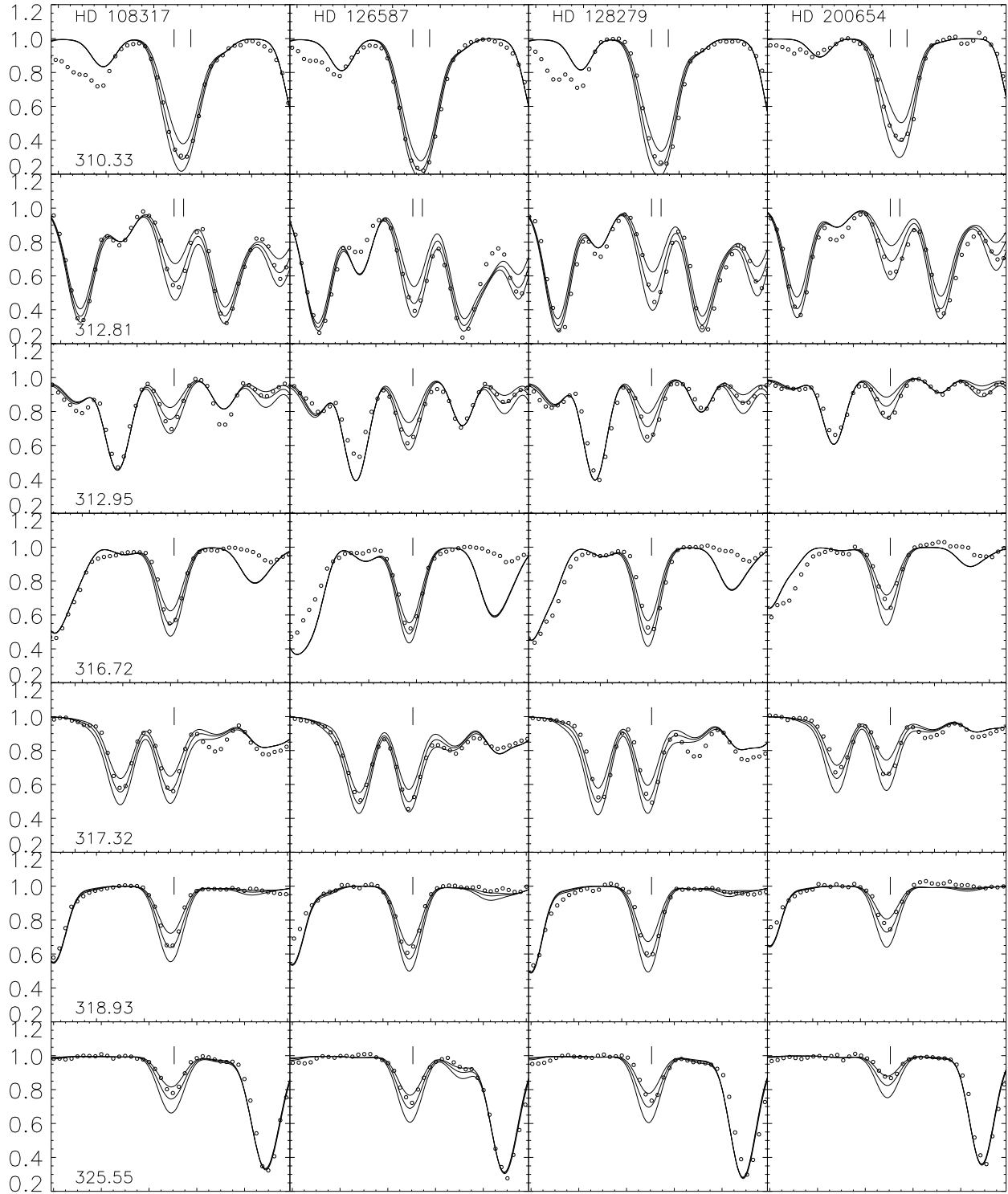}}
 \caption{\label{oohob} Comparison between observed and synthetic spectra
 around the lines in Table\,\ref{ohlist} for 
\object{HD\,108317}, \object{HD\,126587}, \object{HD\,128279} and 
\object{HD\,200654},  with the synthetic spectra computed for three different 
abundances:
 the mean value for each star as listed in Table\,\ref{oh} and these values 
$\pm0.2\dex$.
 The position of the OH lines used 
are marked with vertical bars and their wavelengths are given in the respective
 panel.}  
 
\end{center}
 \end{figure*}

The selection of the OH lines to use was based on spectral synthesis
 for two stars bracketing the metallicity range of our sample, 
\object{HD\,4306} ([Fe/H]=$-2.33$)
 and \object{HD\,274939} ([Fe/H]=$-1.49$).
 Only lines which were essentially
 free of blends and sufficiently strong also in our most metal-poor stars were 
considered.
 The theoretical OH wavelengths, excitation potentials and transition 
probabilities were
 adopted from the line list of \citet{Gillis01}. Many very weak lines which have no impact on
 the theoretical spectrum were culled from the original line list to save time 
in the synthesis,
 which was carried out with the {\sc bsyn} code of the Uppsala synthesis 
package.
 The data for the atomic lines present in this region were taken from VALD 
\citep{Piskunov95} while the necessary data for the few significant CH lines were taken from the 
database of
 \citet{Kurucz93}.
 The first guess for the element abundances used for the synthesis were the 
same as
 in the employed model atmosphere, i.e. solar abundances scaled to the adopted 
metallicity
 with an $\alpha$-enhancement of 0.4\,dex.
 Individual abundances were adjusted as necessary to obtain the correct 
strengths of lines.
 The theoretical spectra were first convolved with a Gaussian appropriate to the
 instrumental resolving power after which an additional broadening profile, 
taking
 care of macroturbulence, rotation etc. was applied.
 The width of the latter was determined individually for each star by 
optimising the overall
 fit between the synthetic and observed spectra both for atomic and molecular 
lines.
 In the end we identified seven spectral features between 310 and 330\,nm which 
were suitable
 for O abundance determinations in metal-poor
 subgiants and giants near the base of the red giant branch. These are listed in
 Table\,\ref{ohlist} together with data for the lines. Two of the features are 
in fact
 composed of two OH lines so data for nine lines are given.
 The OH 316.717 line is partly blended with a CH line, which, however, does not
 cause any significant problems.
 The O abundance determined from this line is consistent with the values
 coming from the other OH lines.

The derived oxygen abundances from each feature are listed in Table\,\ref{oh}, 
together
 with the mean abundance and standard deviation for each star. Spectral 
synthesis
 for all the seven lines together with the observations are 
displayed in Fig.\,\ref{oohob} for four stars. All lines give consistent 
results, which
 suggests that unidentified blends are not a significant problem. The typical 
line-to-line scatter is only about 0.1\,dex, and 0.12\,dex  in the worst case. The 
observed
 OH lines of interest fall in general between the lower and upper solid lines in
 the figure which 
are the theoretical predictions corresponding to abundance variation of 
$\pm0.2\dex$.
 
The scatter can likely be explained by the finite $S/N$, possible blends and 
slightly
 erroneous $gf$-values. Departures from LTE could perhaps also contribute 
\citep{Hinkle75} although
 very little work has been done on NLTE effects for molecular lines of 
electronic band
 systems.
 One would expect, however, that the differential NLTE effects for different 
OH lines
 of similar excitation potential and line strength in the case of our lines 
should
 be marginal, even in the presence of significant overall departures from LTE in
 the molecular number densities or line-source functions. We note that there is 
no
 clear tendency in LTE abundance differences between results from OH lines with 
different
 excitation energies or lines from different vibrational bands.

We have investigated the sensitivity of the derived oxygen abundances to the 
adopted
 stellar parameters and different continuum location in the observed spectra 
within reasonable
 limits. For the purpose, theoretical curve-of-growths were computed for the 
stellar parameters of our programme stars.
 The resulting behaviour is complex as the sensitivity depends both on the 
stellar parameters and the oxygen abundances. These values were perturbed 
with $\Delta T_{\rm{eff}}=100$,
 $\Delta \log g$ according to the errors quoted in Table\,\ref{sp}, and 
$\Delta$[Fe/H]$=0.1$.
 The estimated abundance errors are given in Table\,\ref{senOH}.
 The derived O abundance is quite sensitive to the adopted \teff\ and \logg . 
An increase
 in \teff\ by 100\,K would lead to about 0.2\,dex higher O abundances while a 
0.2\,dex increase
 in \logg\ roughly translates to an decrease in abundance by 0.07\,dex.
 The sensitivity to the metallicity of the model atmosphere is minor.
 It should be noted that the errors due to \teff\ and \logg\ goes
 in the opposite direction to those of \feii , which
 unfortunately makes the derived \ofe\ even more uncertain.
 We have studied the sensitivity of the adopted spectrum normalisation in 
\object{HD\,45282}, which has
 one of the highest \feh\ in our sample and thus should be the most sensitive 
to the normalisation.
 An alternative normalisation by 8\%, which is on the order of what is still 
acceptable,
 leads to a 0.1\,dex difference in the derived O abundance for this star.
 Clearly, this error is small compared with those associated with the stellar 
parameters.
 
The final total errors in the OH-based \ofe\ abundances were estimated for 
each star individually by adding in quadrature those from the stellar 
parameter uncertainties and the error in the mean value of
 the results of the OH feature fits. Typical resulting values are $\pm 0.3$\,dex. 

\begin{table}
 \begin{center}
 \caption{\label{ohlist} Wavelengths of the selected OH lines from the (A-X) 
electronic system
 for oxygen abundance determination, together with their corresponding 
vibrational numbers,
 excitation potential and $\log{gf}$ value.}
 \begin{tabular}{cccc}
 \hline
 \hline
 $\lambda$ & $\nu^{\prime}$-$\nu^{\prime\prime}$& $\chi$ & $\log{gf}$ \\
 {[nm]}& & [eV] \\
 \hline
 310.327 & 0 0 & 0.053 & $-2.799$ \\
 310.334 & 0 0 & 0.053 & $-2.355$ \\ 
312.806 & 1 1 & 0.541 & $-2.581$ \\
 312.810 & 0 0 & 0.210 & $-2.969$ \\
 312.954 & 1 1 & 0.516 & $-2.679$ \\
 316.717 & 0 0 & 1.109 & $-1.544$ \\
 317.320 & 1 1 & 0.842 & $-1.888$ \\
 318.931 & 1 1 & 1.032 & $-1.843$ \\
 325.549 & 0 0 & 1.300 & $-1.811$ \\
 \hline
 \end{tabular}
 \end{center}
 \end{table}
 
\begin{table*}
 \begin{center}
 \caption{\label{oh} The derived 1D LTE oxygen abundances from the OH lines 
listed in Table\,\ref{ohlist}.
 The last column lists the derived mean abundance and its standard deviation 
(s.d.).}
 \begin{tabular}{lcccccccccccc}
 \hline
 \hline
 Star &\multicolumn{8}{c}{$\log{\epsilon}_{\mathrm{O}}$} & $\log{\epsilon}_{\mathrm{O}}\pm{s.d.}$ \\
   & $310.327+$ & $312.806+$ & 312.954 & 316.717 & 317.320 & 318.931 & 325.549  
 & \\
   & 310.334 & 312.810 & & & & & & & & \\
 \hline
 \object{HD\,4306}   & 6.65 & 6.73 & 6.80 & 6.63 & 6.68 & 6.68 & 6.58 &  & $   6.68 \pm  0.07 $ \\
 \object{HD\,26169}  & 6.90 & 7.00 & 7.00 & 6.90 & 6.97 & 6.89 & 6.82 &  & $   6.93 \pm  0.07 $ \\
 \object{HD\,27928}  & 6.96 & 7.06 & 7.08 & 6.90 & 6.95 & 6.90 & 6.86 &  & $   6.96 \pm  0.08 $ \\
 \object{HD\,45282}  & 7.85 & 7.90 & 7.98 & 7.90 & 8.00 & 7.90 & 7.77 &  & $   7.90 \pm  0.08 $ \\
 \object{HD\,108317} & 7.20 & 7.30 & 7.33 & 7.25 & 7.25 & 7.22 & 7.15 &  & $   7.24 \pm  0.06 $ \\
 \object{HD\,126587} & 6.30 & 6.38 & 6.40 & 6.25 & 6.40 & 6.20 & 6.22 &  & $   6.31 \pm  0.09 $ \\
 \object{HD\,128279} & 7.30 & 7.40 & 7.45 & 7.25 & 7.35 & 7.25 & 7.21 &  & $   7.32 \pm  0.09 $ \\
 \object{HD\,200654} & 6.82 & 6.97 & 7.02 & 6.82 & 6.82 & 6.82 & 6.70 &  & $   6.85 \pm  0.11 $ \\
 \object{HD\,218857} & 7.07 & 7.12 & 7.22 & 7.07 & 7.12 & 7.12 & 7.05 &  & $   7.11 \pm  0.06 $ \\
 \object{HD\,274939} & 7.51 & 7.65 & 7.70 & 7.55 & 7.65 & 7.60 & 7.56 &  & $   7.60 \pm  0.07 $ \\
 \bd                 & 6.67 & 6.82 & 6.85 & 6.78 & 6.75 & 6.75 & 6.75 &   & $   6.77 \pm  0.06 $ \\
 \cda                & 6.75 & 6.95 & 6.95 & 6.73 & 6.78 & 6.73 & 6.72 &  & $   6.80 \pm  0.10 $ \\
 \cdb                & 6.33 & 6.41 & 6.33 & 6.20 & 6.27 & 6.16 & 6.05 &  & $   6.25 \pm  0.12 $ \\
 \hline
 \end{tabular}
 \end{center}
 \end{table*}
   
\begin{table}
 \begin{center}
 \caption{\label{senOH} Errors in abundances due to
 uncertainties in effective temperature of 100\,K, $\Delta T_{\rm{eff}}$,
 in \logg\ as given in Table\,\ref{sp}, $\Delta \log{g}$ , and in metallicity
 of 0.1\,dex, $\Delta$[Fe/H].}
 \begin{tabular}{lccc}
 \hline
 \hline
$Star$ & \multicolumn{3}{c}{$\Delta \log{\epsilon}_{\mathrm{O}}$} \\
  &$\Delta T_{\rm{eff}}$ & $\Delta \log{g}$ & $\Delta$[Fe/H] \\
 \hline
 \object{HD\,4306}  &  0.19  & $-0.06$  &  0.04 \\
 \object{HD\,26169} &  0.21  & $-0.07$  &  0.03 \\
 \object{HD\,27928} &  0.20  & $-0.12$  &  0.03 \\
 \object{HD\,45282} &  0.17  & $-0.03$  &  0.05 \\
 \object{HD\,108317}&  0.22  & $-0.07$  &  0.02 \\
 \object{HD\,126587}&  0.25  & $-0.19$  &  0.02 \\
 \object{HD\,128279}&  0.20  & $-0.07$  &  0.02 \\
 \object{HD\,200654}&  0.17  & $-0.12$  &  0.01 \\
 \object{HD\,218857}&  0.17  & $-0.07$  &  0.06 \\
 \object{HD\,274939}&  0.15  & $-0.06$  &  0.06 \\
 \bd                &  0.20  & $-0.09$  &  0.04 \\
 \cda               &  0.19  & $-0.09$  &  0.04 \\
 \cdb               &  0.23  & $-0.16$  &  0.01 \\
\hline
 \end{tabular}
 \end{center}
 \end{table}

 
\section{Discussion}
 \label{discussion}
 
\subsection{Comparison between the three different indicators}

Our comparison between the results from the three oxygen abundance indicators is
 made primarily in terms of [O/Fe] which is a key parameter in studies of 
the chemical evolution of oxygen in the Galaxy. In this way, effects of errors 
in the stellar parameters are reduced in cases when derived oxygen and iron 
abundances respond in similar ways to changes in the parameters, i.e. for the 
\oifor\ and \oi\ .
 
\begin{table*}
 \begin{center}
 \renewcommand{\tabcolsep}{3pt}
 \caption{\label{ofet} [O/Fe] values and their uncertainties as determined from 
\ion{Fe}{ii} lines and the three oxygen indicators:
 {\oifor}, {\oi} IR and OH UV lines. The values derived from {\oi} IR 
lines are based on NLTE calculations.}
 \begin{tabular}{lcccccccccccccccccc}
 \hline
 \hline
 Star&[O/Fe]& & & & &[O/Fe]& & & & & [O/Fe]& & & & \\ 
 &[\ion{O}{i}] & $\Delta T_{\rm{eff}}$ & $\Delta \log g$ & $\Delta$[Fe/H] & $\Delta{W_\lambda}$ &\ion{O}{i} & $\Delta T_{\rm{eff}}$ & $\Delta \log g$ & $\Delta$[Fe/H] & $\Delta{W_\lambda}$  & OH & $\Delta T_{\rm{eff}}$ & $\Delta \log g$ & $\Delta$[Fe/H] \\
 
 \hline
 \object{HD\,4306}  & 0.73   &  0.06 &  0.00 &  0.01 &  0.09 &   0.95 & $-0.10$ & $ 0.01$ & $-0.01$ &  0.10  &  0.27  &  0.21  & $-0.16$  &  0.03 \\
 \object{HD\,26169} & 0.59   &  0.06 &  0.00 &  0.01 &  0.07 &   0.85 & $-0.10$ & $-0.00$ & $-0.01$ &  0.09  &  0.47  &  0.21  & $-0.17$  &  0.02 \\
 \object{HD\,27928} & 0.46   &  0.06 & $-0.01$ &  0.01 &  0.09 &   0.62 & $-0.09$ & $ 0.00$ & $-0.01$ &  0.11  &  0.36  &  0.21  & $-0.29$  &  0.02 \\
 \object{HD\,45282} & 0.50   &  0.04 &  0.00 &  0.00 &  0.06 &   0.48 & $-0.10$ & $-0.01$ & $-0.03$ &  0.03  &  0.68  &  0.17  & $-0.08$  &  0.03 \\
 \object{HD\,108317}& 0.74   &  0.07 &  0.00 &  0.01 &  0.09 &   0.68 & $-0.09$ & $ 0.01$ & $-0.01$ &  0.07  &  0.75  &  0.21  & $-0.14$  &  0.02 \\
 \object{HD\,126587}& 0.45   &  0.08 & $-0.01$ &  0.01 &  0.11 &   1.10 & $-0.11$ & $ 0.04$ & $ 0.00$ &  0.20  &  0.44  &  0.26  & $-0.37$  &  0.02 \\
 \object{HD\,128279}& 0.41   &  0.07 &  0.00 &  0.01 &  0.22 &   0.62 & $-0.09$ & $ 0.01$ & $-0.00$ &  0.08  &  0.77  &  0.21  & $-0.13$  &  0.02 \\
 \object{HD\,200654}&$<0.74$ &  .... &  .... &  .... &   ....&   0.62 & $-0.10$ & $ 0.02$ & $-0.01$ &  0.22  &  0.84  &  0.21  & $-0.24$  &  0.01 \\
 \object{HD\,218857}& 0.38   &  0.06 &  0.01 &  0.01 &  0.06 &   0.61 & $-0.09$ & $-0.01$ & $-0.02$ &  0.06  &  0.16  &  0.19  & $-0.21$  &  0.04 \\
 \object{HD\,274939}& 0.51   &  0.03 & $-0.02$ & $-0.01$ &  0.03 &   0.42 & $-0.09$ & $-0.01$ & $-0.02$ &  0.05  &  0.35  &  0.17  & $-0.20$  &  0.04 \\
 \bd&  0.54   &  0.06 &  0.00 &  0.01 &  0.09 &   0.87 & $-0.09$ &  0.01 & $-0.01$ &  0.06  &  0.13  &  0.20  & $-0.24$  &  0.03 \\
 \cda&  ....   &  .... &  .... &  .... &  .... &   0.69 & $-0.09$ &  0.01 & $-0.01$ &  0.11  &  0.30  &  0.20  & $-0.23$  &  0.04 \\
 \cdb& $<0.65$ &  .... &  .... &  .... &  .... & $<0.88$&  .... &  .... &  .... &  ....  &  0.52  &  0.22  & $-0.34$  &  0.01 \\
 \hline
 \end{tabular}
 \end{center}
 \end{table*}

\begin{figure}
 \begin{center}
 \resizebox{\hsize}{!}{\rotatebox{0}{\includegraphics{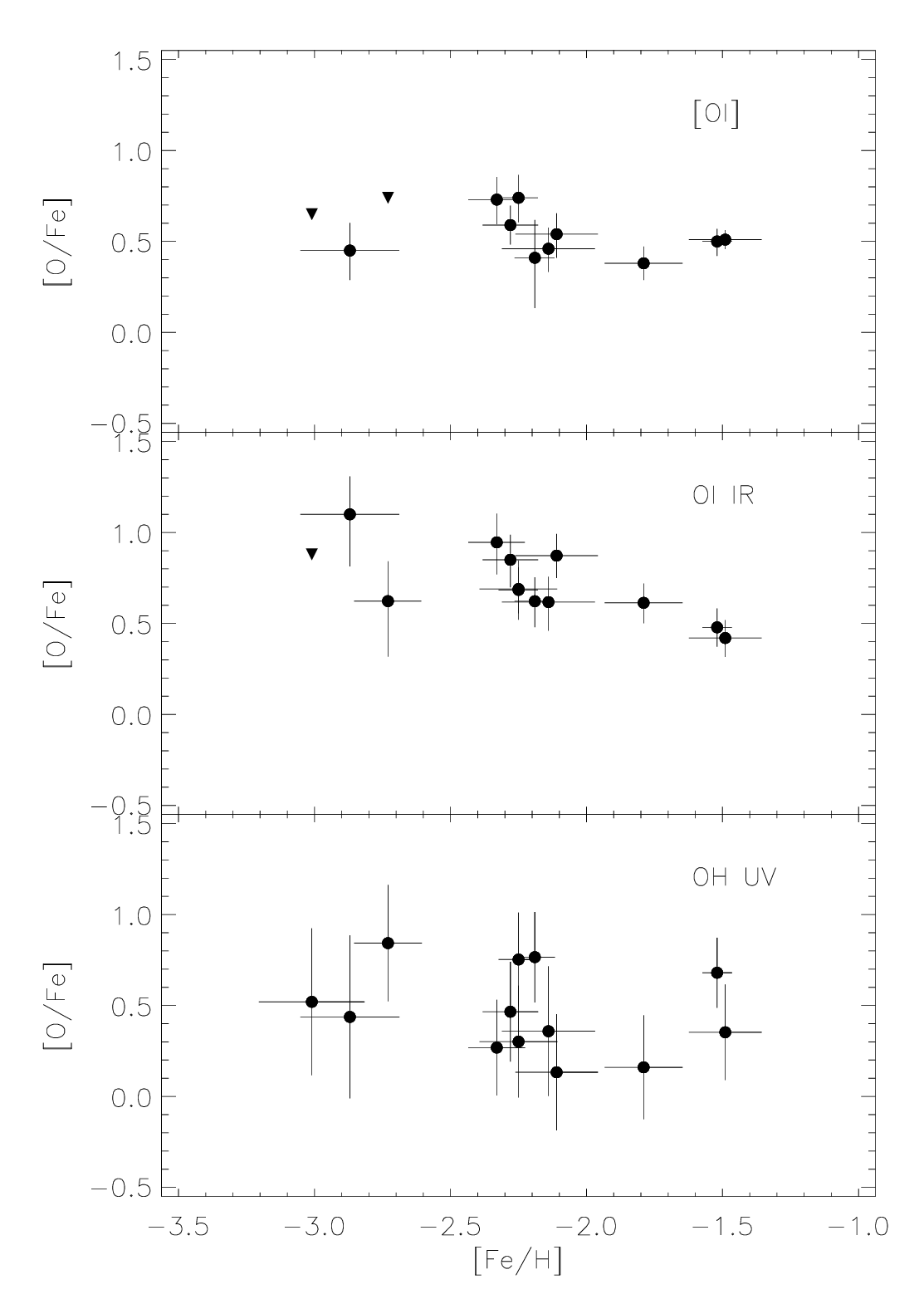}}}
 \caption{\label{ofec}[O/Fe] estimates as a function of metallicity. 
Estimates based on different indicators are shown separately: \oifor\ 630.03\,nm (top panel), \oi\ 777.1-5\,nm (middle panel) and OH UV 310.0\,nm (bottom panel).}
 \end{center}
 \end{figure}

Values of [O/Fe] and their uncertainties are summarised in Table\,\ref{ofet}. 
In Fig.\,\ref{ofec} and for each indicator separately, these values are shown 
as a function of metallicity. The values derived from OH (bottom panel) are by far the most uncertain with typical error bars in [O/Fe] of the order of 
$0.30\dex$. About two times smaller than this are [Fe/H] and [O/Fe] error bars 
associated to the forbidden and permitted {\oi} lines. The effective 
temperature plays
 a major role in the determination of accurate 
[O/Fe] values independently of the indicator used. Gravity only plays a 
significant role when oxygen abundances are derived from the OH UV lines. Measurement 
errors in equivalent widths also contribute to the [O/Fe] total error.
 
For many decades, the accuracy of 1D-LTE [O/Fe] values based on {\oifor} has 
been
 limited by the quality of the observed stellar spectra. With the era of new 
astronomical instruments, this quality has improved very considerably such that
 measurements based on the weak forbidden line, when not too weak,
 now probably give the most reliable estimates 
of [O/Fe] in metal-poor stars. Here, these estimates are taken as reference
 for the comparison between the indicators.
 
\begin{figure}
 \begin{center}
 \resizebox{\hsize}{!}{\rotatebox{0}{\includegraphics{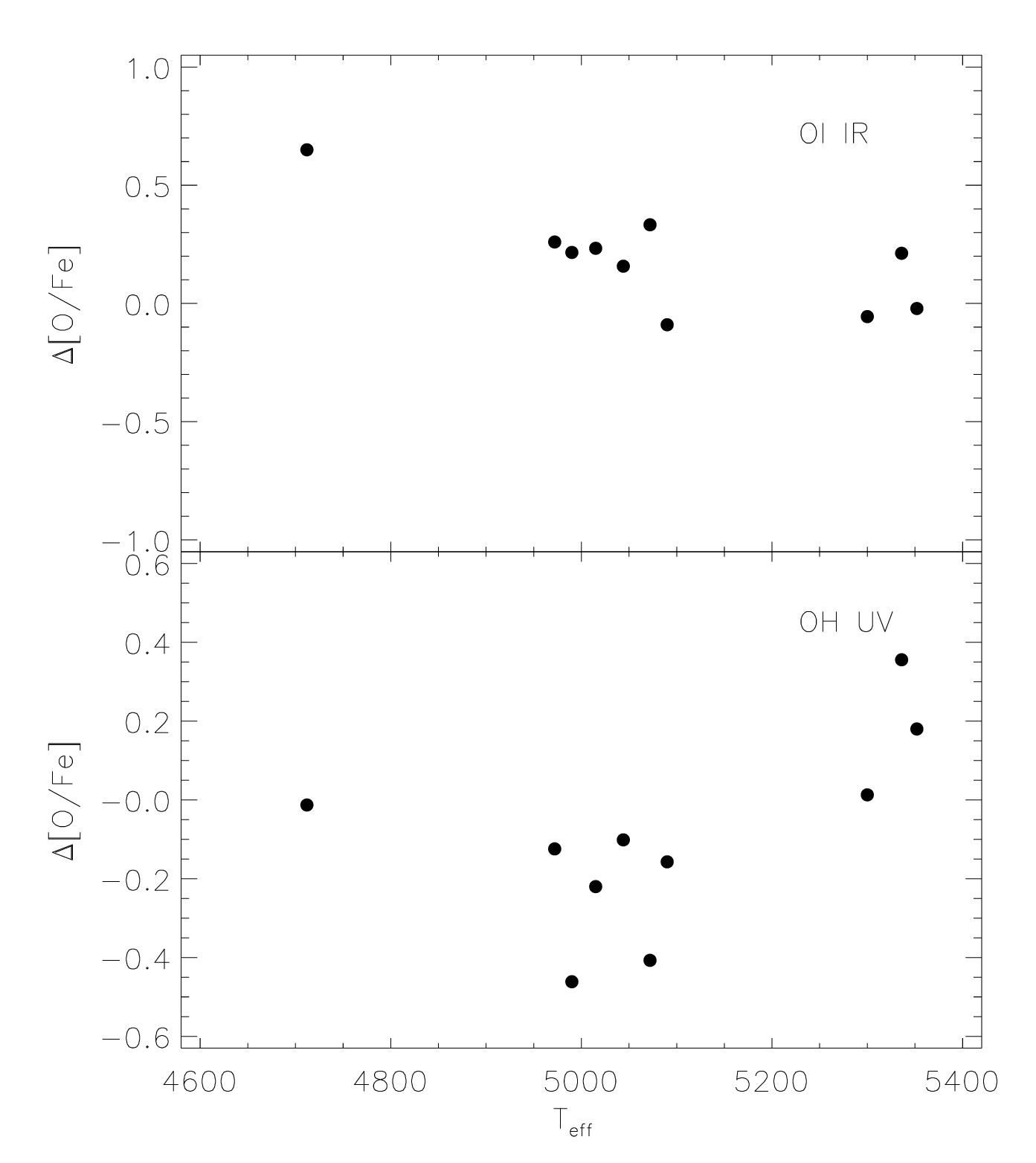}}}
 \caption{\label{ofeteff} Departures of {\oi}-based (top panel) and 
OH-based (bottom panel) abundance estimates from [{\oi}]-based abundances as 
a function of \teff.}
 \end{center}
 \end{figure}
 
According to Table \ref{ofet}, [O/Fe] values based on {\oi} permitted lines 
(including NLTE corrections) are on average $0.19\pm0.07\dex$ higher than the 
reference values, while values based on OH UV lines are $0.09\pm0.08\dex$ 
lower. The magnitudes of the [O/Fe] uncertainties for the individual stars 
are of the same order or even higher than these values, except the 
uncertainties of the {\oifor}-based [O/Fe]. Notice that in the total error 
in [O/Fe], uncertainties associated with the analysis of the \ion{Fe}{ii} 
lines should be included. These uncertainties, however, affect estimates 
based on the different criteria identically, and thus 
do not have any impact on the differences. Typical [O/Fe] 
error bars are of the order of 0.12, 0.15 and $0.3\dex$ for the estimates 
based on {\oifor}, {\oi} and OH UV lines, respectively. 

If our {\teff}-scale turns out to be too cool, then both {\oifor} and 
{\oi}-based estimates
 are expected to need modification. An increase of the temperatures by 100\,K
 will bring the two different estimates together because the two indicators 
respond
 in opposite directions to these changes, with shifts in [O/Fe] of $0.06~\dex$ 
and $-0.1\dex$,
 respectively. Another way of explaining the differences would be to advocate 
systematic
 errors in the equivalent widths. E.g., if we assume that 
the forbidden and the permitted lines are too strong and too weak, respectively,
 by their typical equivalent-width uncertainties as estimated above, the 
{\oifor}-based [O/Fe] values should be adjusted upwards 
and the {\oi}-based [O/Fe] values decreased 
by about the same amount, $\sim0.1\dex$, and thus reach agreement. A too 
cool {\teff}-scale can 
also lead to underestimated [O/Fe] values when based on abundance 
determinations
 from OH but by different amounts than for {\oifor}. Estimates 
based on the molecular lines increase faster with effective temperature than 
estimates
 based on the forbidden line ($0.20$ versus $0.06\dex$) and hence the two 
estimates get closer
 to each other with a hotter temperature scale. The mean difference in [O/Fe] 
between
 values based on \oifor\ and OH, respectively, is lower than the [O/Fe] uncertainties 
of about $0.20\dex$ due to gravity errors in the analysis of OH lines.
 
\begin{figure}
 \begin{center}
 \resizebox{\hsize}{!}{\rotatebox{0}{\includegraphics{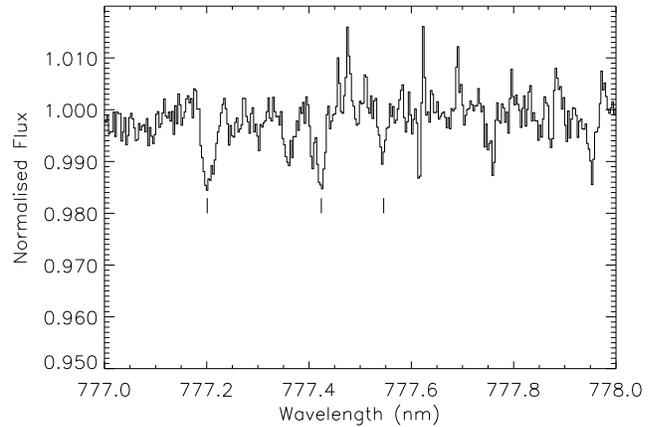}}}
 \caption{\label{giant} UVES-spectra of \object{HD\,126587} around the 
{\oi} IR triplet 777.1-5\,nm lines.
 Vertical bars point out the expected positions of the three lines.}
 \end{center}
 \end{figure}
 
Unfortunately, the standard deviations in the differences are quite 
significant in comparison with
 the mean differences themselves. There are a few stars
 for which the differences are much higher than the mean value. 
Fig.\,\ref{ofeteff} even demonstrates possible 
relations between the differences and the effective 
temperature; higher effective temperature may suggest a lower difference 
between abundance ratios from the triplet lines and the forbidden line, 
respectively.
 The trend for the abundance difference from the OH lines and the forbidden 
line is inverted with respect to this; only the coolest star, \object{HD\,126587}, seems to break such a trend since 
it shows a good agreement between the two indicators. Unfortunately, this is not the 
case for the triplet lines in the spectrum of the same star; this giant has the 
highest discrepancy ($0.65\dex$) between abundances from those and from the forbidden 
line. \object{HD\,126587} has also high observational errors in \oi-based [O/H], $0.24\dex$; 
these uncertainties together with uncertainties in effective temperature are, 
however, not high enough to account for the observed differences between 
results from the two different criteria. The observed equivalent widths of 
the {\oi} IR triplet lines of this star are so low that ripples or fringing 
in the observed spectra might have disturbed the measurements, see Fig.\,\ref{giant}. 
In the literature we can find other cases with high discrepancies between the two 
indicators. E.g., according to \citet{Barbuy03}, the [O/Fe] estimates for the 
metal-poor giant \object{HD\,122563} differ by 0.5~dex.
 
Differences between the OH-based estimates as compared with those of the forbidden line
 are compatible with uncertainties in the corresponding [O/Fe] estimates if 
they are smaller than about $0.3\dex$. Typically, an increase in effective 
temperature by 100\,K will increase the estimates by $0.14\dex$ and reduce the 
differences relative to the estimates from the forbidden line, as will
 a decrease in gravity by an amount consistent with the expected errors in this quantity
 which will increase the estimate by about $0.2\dex$. Also, a decrease of 
the equivalent width of the forbidden line by its maximum error, 0.05\,pm, 
would bring down the [O/Fe] estimates from that line by $\sim0.1\dex$. As 
a result of all this, the differences between the different [O/Fe] estimates of 
about $0.3\dex$ would be reduced to zero. There are three data points, representing 
\object{HD\,4306}, \object{HD\,128279} 
and \object{BD$-01\degr 2582$} in the bottom panel in Fig.\,\ref{ofeteff}, which 
depart more than one would expect from the errors in parameters and observations. 
The differences plotted are negative for two of them 
while the third one has positive differences (\object{HD\,128279}).

We remind the reader that our comparison between \ion{Fe}{i}- and \ion{Fe}{ii}-based 
iron abundances for \object{HD128279} may suggest that our derived effective temperature 
has been overestimated (cf Sect.\,\ref{metallicity}). If this were the case, then the [O/Fe] 
differences plotted 
in Fig.~\ref{ofeteff} as well as the scatter in the [O/Fe]-vs-[Fe/H] trend would be notably reduced 
(a decrease of 100~K in \teff\ would lower the OH- and {\oifor}-based [O/Fe] values by 0.17~dex 
and 0.04~dex respectively). The case of \object{HD\,4306} could be very similar, in the sense 
that for this star the Hipparcos-based gravity may be too large (see Sect.\,\ref{metallicity}). With the 
newly suggested value of $\log g=2.5$, the OH-based [O/Fe] ratios would significantly increase 
(from 0.27 to 0.6), strongly reducing the abundance difference between the two oxygen indicators.
   
{{\bd}} is claimed to 
be a CH binary star \citep{Carney03} which may explain its departure.
 If its carbon abundance is very high then most of its 
oxygen may be in the form of CO and little in the form of OH in its outer 
atmospheric layers; thus a higher oxygen abundance may be required to fit its 
observed OH lines. However, the [O/Fe] values
 based on the forbidden and permitted lines are also in conflict for this star. 
In contrast to this star, with the worse agreement between the abundances from the
 three indicators, there is \object{HD\,108317} with the best agreement: 
[O/Fe] = 0.74 ({\oifor}), 0.68 ({\oi}) and  0.75 (OH). 
 
We have explored whether a change of the \teff\ scale such that an assumed 
shift of {\teff}, which in turn would also be a function of \teff, could provide 
an explanation for the trend found for the abundance differences. We did not find 
any obvious such possibility. We noted, however, that the departure of our 
Hipparcos-based gravity estimates from the isochrone-based estimates depends 
on the value assumed for \teff. The assumed gravity values of the coolest stars 
are much higher than the values based on isochrones, while the values for the 
warmest stars are much lower.
 
\subsection{\label{systematic} Possible systematic errors: NLTE and 3D effects}
 
Up to this point, we have discussed the differences between the oxygen 
abundances derived from the
 different abundance criteria in terms of errors in stellar parameters 
and observations. However we suspect that our 1D-LTE abundance estimates 
are affected by systematic errors.  
 
The analysis of the oxygen IR triplet lines was carried out with consideration 
of
 NLTE effects, and due to this the differences between the results and those of 
the {\oifor} line analysis were reduced. The latter is 
expected to be immune to NLTE effects \citep{Kiselman93}. The 
[O/Fe] estimates may still be subject to other systematic errors, e.g.
 caused by thermal inhomogeneities in the stellar atmospheres.
 At least for main-sequence stars, the 
agreement between the forbidden and permitted lines does not seem to improve 
with
 the use of 3D hydrodynamical model atmospheres, sooner the opposite seems to 
occur \citep{Nissen02}. 
Abundance corrections for oxygen in the 3D model atmospheres for dwarfs of 
low metallicities
 are negative \citep{Nissen02} and they tend to be more important
 for [O/Fe] estimates from {\oifor} than those from the {\oi} triplet. There 
are some
 indications that this may be the case also for subgiants 
\citep{Nissen02,Shchukina04}. 3D calculations for main-sequence stars also suggest that the 3D 
[O/Fe]
 corrections are more significant at low metallicities. From Table\,\ref{ofet} 
one can see that the most-metal poor stars indeed tend to have the greatest 
discrepancies between the results from the different criteria.  

In main-sequence stars the 
OH lines are more affected by 3D effects than the forbidden line, leading to 
enhanced OH lines, i.e. smaller abundances. These effects depend mainly on 
[Fe/H] and to a lesser degree on effective temperature; 3D corrections to
 [O/Fe] decrease with decreasing effective temperature and increasing 
metallicity \citep{Asplund01}. The tendency in 
the bottom panel of Fig.\,\ref{ofeteff} is consistent with these computations. 
However, we do not trace any dependence on 
stellar metallicity although 3D-effects on [O/Fe] from {\oifor} as well
 as from OH are both expected to depend on metallicity 
\citep{Asplund01,Nissen02}.
 
Schematic NLTE calculations for OH UV lines suggest that departures from 
LTE
 line formation occur mainly because the radiation fields and source functions
 are higher than the local Planck function 
\citep{Asplund01}. Therefore, one would expect that NLTE abundance 
corrections
 for the lines in Table\,\ref{ohlist} should be positive (i.e. increase the 
abundances)
 and may depend on effective temperature 
as well as on metallicity. This may be one reason for the discrepancies 
between
 results based on these lines and the {\oifor} line. Typical values for the 
corrections
 in dwarfs are of the order of $0.2\dex$ \citep{Asplund01} and there is 
evidence for a
 dependency on metallicity: the corrections become smaller as the metallicity 
increases. However, these results are based on the two-level 
approach which may be not appropriate for OH lines, thus NLTE effects 
should be investigated further. A full NLTE treatment is expected to increase 
the value of [O/Fe] which 
will compete with the 3D-effects that will decrease [O/Fe].
 
Once 3D-hydrodynamical model atmospheres are available for subgiants of Pop II,
 3D-effects on our [O/Fe] estimates should be investigated.

\subsection{The galactic evolution of oxygen -- comparisons to other studies}
 
We have discussed the discrepancies between the results from the different 
indicators for the oxygen abundance and suggest that our (1D-LTE) results 
based on the {\oifor} 630.03\,nm line are to be preferred. We shall now 
comment on the galactic evolution of oxygen according to these results. 
In recent years, work done especially based on the forbidden line has 
suggested a trend in [O/Fe]-vs-[Fe/H] that is flat at low metallicities 
while several studies based on OH UV electronic transitions 
\citep{Israelian98,Israelian01,Boesgaard99} suggest a linear trend, 
with [O/Fe] gradually rising to high levels at the lowest metallicities. 
 
Among our results some data points around $-2.0$ disturb the picture of 
a possible flat trend in [O/Fe]-vs-[Fe/H] at a level of 0.55 (see top 
panel in Fig.\,\ref{ofec}). The scatter in [O/Fe] around that metallicity is 
barely significant, $\sim0.14$ (s.d.). [O/Fe] ranges from 0.41 to 0.74; the 
values departing most from the mean correspond to \object{HD\,4306} and 
\object{HD\,108317}. While the last star is among those showing the best 
agreement among abundances from the three indicators, \object{HD\,4306} is 
one where the indicators disagree severely. \object{HD\,27928} and 
\object{BD$-01\degr 2582$}, with low [O/Fe] values, are confirmed and claimed 
binaries \citep{Carney03}, respectively, so their abundances may be contaminated.
 Note, however, that \object{HD\,108317} may be a binary, according to 
\citet{Carney03}. \object{HD\,128279} has a low [O/Fe] value, but is another star
 with poor agreement among the indicators. On the other hand, a linear fit to 
the data points in the top panel of Fig.\,\ref{ofec} is difficult to establish, 
especially due to the scatter at low metallicity. We note that our result for 
the metal-poor giant \object{HD\,126587}, with its relatively low [O/Fe] 
($\sim0.45$), has a great effect on attempts to judge whether 
there is a "flat" tendency or a gradual increase in [O/Fe] when proceeding 
towards low metallicities. The result of a linear fit to the data points is 
$\mathrm{[O/Fe]}=-0.09(\pm0.08)\mathrm{[Fe/H]}+0.36(\pm0.15)$. If instead, 
metallicities based on the \ion{Fe}{i} are assumed, this 
trend becomes steeper. This is due to the dependence on metallicity of the iron abundance 
differences discussed in Sect.\,\ref{metallicity}. 

\begin{figure}
 \begin{center}
 \resizebox{\hsize}{!}{\rotatebox{0}{\includegraphics{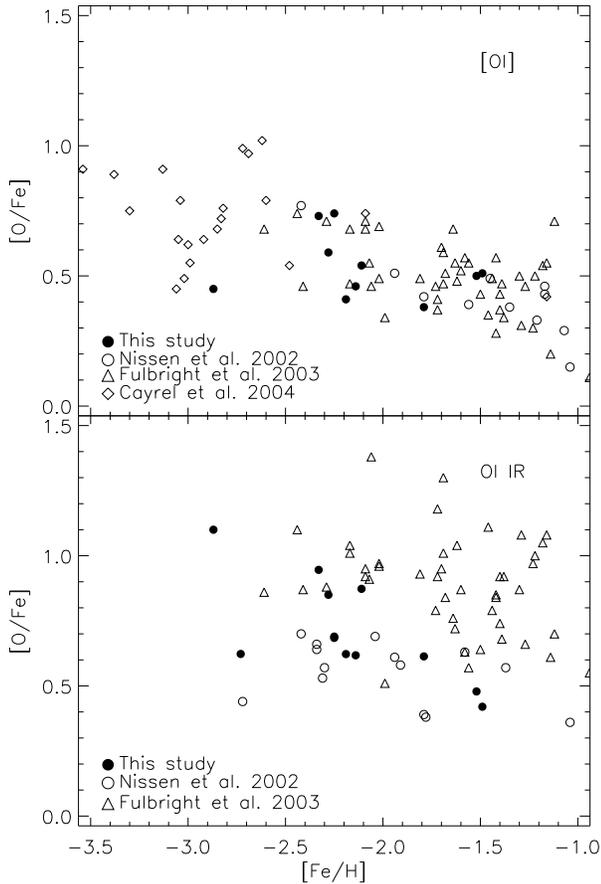}}}
 \caption{\label{ofec2} Comparison of [O/Fe] values in Table\,\ref{ofet} with 
results from:
 \citet{Nissen02}, \citet{Fulbright03} and \citet{Cayrel04}. Results based on 
the \oifor\ line
 are shown in the top panel while results based on the {\oi} triplet are shown 
in the bottom panel.
 A consistent value for the solar oxygen abundance, $8.74\dex$, was adopted for 
all
 the data points.}
 \end{center}
 \end{figure}

We have chosen to compare our results with four studies in which the 
metallicity ranges overlap with ours: \citet{Israelian98,Nissen02,Fulbright03,Cayrel04}.
 The results in \citet{Nissen02} are easy to compare with because we 
have used the same method and the same tools. The methods used by 
\citet{Fulbright03} and
 \citet{Cayrel04} are also very similar to ours although the tools are partly 
different.
 The method utilised by \citet{Israelian98} is somewhat deviating 
from ours. No
 attempt was made to homogenize the scales of stellar parameters in the 
comparison with
 them. In the \citet{Fulbright03} case, 
we could use the option to compare to abundance results based on the Alonso et 
al. temperature scale.
 [O/Fe] values plotted in Fig.\,\ref{ofec2} were calculated from the values of 
[Fe/H] and \logo \, as reported
 in the different papers after corrections of oxygen abundances for differences 
in the assumed $\log{gf}$-values.
 For the solar oxygen abundance, we took $8.74\,\dex$, see Sect.\,\ref{sun}.

The different studies cover different stellar evolutionary phases. While
 \citet{Israelian98} and \citet{Nissen02} mainly study main-sequence stars, 
\citet{Fulbright03} and \citet{Cayrel04} analyse subgiants or giants. Our 
study goes to lower metallicities than the studies of Nissen et al. and 
Fulbright et al. but not as low as Cayrel et al. (Fig.\,\ref{ofec2}). In general, 
our [O/Fe] values, whether based on the forbidden or the triplet lines, follow 
the same tendencies with [Fe/H] as in the other studies. For oxygen abundances 
from the forbidden line, see top panel in Fig.\,\ref{ofec2}, there is no clear 
distinction between our and the rest of the data points. For the triplet line results
 (bottom panel) there is a systematic difference between, on one hand our 
results and those of Nissen et al., relative to
 those of Fulbright et al. on the other; a differentiation which seems to occur 
between dwarfs and giants.
 In that plot, 
most of our subgiants lie close to the dwarfs from Nissen et al., especially 
the not very cool ones.
 A few of the cooler stars get instead close in [O/Fe] to the giants of Fulbright 
et al.
 
\begin{figure}
 \begin{center}
 \resizebox{\hsize}{!}{\rotatebox{0}{\includegraphics{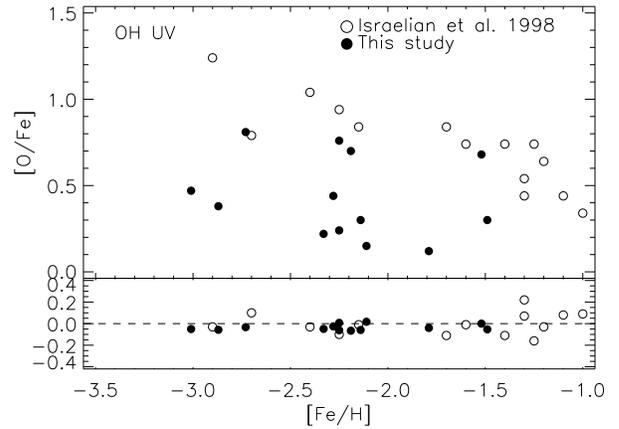}}}
 \caption{\label{ofeOH} Comparison between our results and those of 
\citet{Israelian98} as regards [O/Fe] values based on the OH line at 316.7\,nm. The bottom panel 
shows how the [O/Fe] values based on the single OH line at 316.7\,nm deviate from 
the mean values for all OH lines used in the study. The same solar oxygen abundance, 
8.74, was used for all data points.}
 \end{center}
 \end{figure}
 
We have three stars in common with \citet{Fulbright03}: \object{HD\,45282}, 
\object{HD\,108317} and
 \object{HD\,218857}. These authors derived oxygen abundances from both the 
permitted and the forbidden oxygen
 lines using two different stellar parameter scales. We have compared our 
results with their results
 for the Alonso et al. scale with consideration of the differences in the 
assumed stellar parameters, and find that 
our and their abundances agree to within 0.1\,dex. We note that they use 
Kurucz model atmospheres and that their oxygen abundance estimates are based 
on the use of both forbidden lines.
 
In order to compare our results from OH with the linear fit suggested by 
\citet{Israelian98}, it is necessary to 
displace their fit by 0.2\,dex along the vertical axis to account for the 
difference in the adopted
 solar oxygen abundance. Our results lie below theirs. 
This could be due to the use of a different set of OH UV 
lines or due to the use of different oscillator strength values. 
These possibilities have been explored by a comparison 
based on the only OH line in common (316.7~nm). The oxygen abundances were 
corrected for the differences
 in the $\log{gf}$-value ($-1.694$ versus $-1.544$), and plotted in 
Fig.\,\ref{ofeOH}. Obviously, our data
 points lie significantly below those of Israelian et al.; only the subgiants 
with the highest \teff\ may be close. We remind the reader that the 
effective temperature for two of these stars (\object{HD\,45282} and \object{HD\,128279}) 
may have been overestimated and that if a decrease of 100~K in their temperatures 
(see Sect.~\ref{metallicity}) is taken into account the data points corresponding to 
these stars will move down in the figure and consequently away from the ones of 
Israelian et al. However, the decrease in gravity suggested for other two stars of 
our sample (\object{HD\,4306} and \object{\cda}) in Sect.\,\ref{metallicity} would 
instead move up these two objects. In conclusion, although 
star-to-star differences may change if the above mentioned 
changes in \teff\ and \logg\ respectively are applied, the mean difference between our 
abundances and Israelian et al. is almost unaffected.

 The bottom panel in Fig.\,\ref{ofeOH} shows how the [O/Fe] values based on the single line depart from the mean values for all OH lines used in each study. These departures 
are quite small compared with the differences between us and Israelian et al. Those 
differences are probably due to the fact that Israelian et al. analysed stars close to the 
turn-off point. These stars are expected to show greater effects due to 
thermal inhomogeneities
 on the abundances derived from OH lines than our cooler subgiants 
\citep{Asplund01,Nissen02}.
 
In general, the systematic differences in oxygen abundances between dwarfs and 
giants
 found in the present study may most probably be explained 
as results of a change with effective temperature and surface gravity 
of the systematic errors, as was already suggested by line 
computations for OH UV lines in 3D model atmospheres by \citet{Asplund01}.
 
\section{Conclusions}
 
Until 3D hydrodynamic model atmospheres for subgiants are available which will 
make 3D-NLTE line formation computations possible, 1D-LTE analyses based on the 
forbidden 630.03\,nm line of oxygen will probably provide the best estimates of 
stellar [O/Fe] when observable. These values are by far the most accurate, with 
typical errors of the order of $0.10$\,dex. Note although that they may not be 
free of significant systematic errors associated with inhomogeneities in the 
stellar atmospheres. Our best [O/Fe] estimates suggest a
 value at a level of $0.55\pm0.13$ for Pop II subgiants, relatively independent 
on metallicity; the best linear fit to these estimates is 
$\mathrm{[O/Fe]}=-0.09(\pm0.08)\mathrm{[Fe/H]}+0.36(\pm0.15)$. 

Both [O/Fe] values based on the IR triplet lines and on the OH UV lines 
depart from the the {\oifor}-based estimates. The departures from the {\oifor} results are
 more significant for the abundances derived from the triplet lines, on average 
$0.19\pm0.22\dex$ (s.d.), than for the values from the molecular lines, 
$-0.09\pm0.25\dex$ (s.d.). In the case of the triplet lines, the departures 
decrease with increasing effective temperatures while in the case of the OH lines they 
increase, turning from being negative to positive. According 
to line computations for 3D model atmospheres of main-sequence stars, 3D abundance 
corrections are negative and decrease with decreasing effective temperature
 which is not in contradiction with the present observations. In general, 3D effects on 
abundances and departures from LTE depend on stellar parameters and this can be the 
reason for the fact that differences in [O/Fe] between the indicators depend on the 
fundamental stellar parameters.

In order to increase the accuracies of [O/Fe] the $S/N$ in the observations 
should be further improved and the effective temperatures should be determined
 more accurately. Accurate gravities are as well necessary to improve the values 
based on OH UV lines whose typical errors are of the order of $0.3\dex$.

Our results for subgiants follow the trends found by many others for main-sequence 
stars and giants in the last few years with the exception of our OH-based results
 which are significantly lower. 

\begin{acknowledgements} We thank A. Goldman for providing us with 
the $\log{gf}$ values for OH lines. We are grateful to K. Eriksson and N. Piskunov 
for fruitful discussions. An important part of this work was carried out during 
the visit of the first author to ESO and she much appreciate 
their hospitality and financial support. The work had been supported 
by the Nordic Optical Telescope and the Swedish Research Council. This publication 
makes use of data products from the Two Micron All Sky Survey, which is a joint project 
of the University of Massachusetts and the Infrared Processing and Analysis 
Center/California Institute of Technology, funded by the National Aeronautics and 
Space Administration and the National Science Foundation.
\end{acknowledgements}
 
\bibliographystyle{aa}
 \bibliography{ref}
 \end{document}